\documentclass[11pt,tightenlines,eqsecnum,floats,aps,amsmath,amssymb,nofootinbib,prd,shownopacs,floatfix]{revtex4}
\usepackage{epsf}
\pdfoutput=1
\usepackage{graphicx, wrapfig}
\usepackage{amssymb}
\usepackage{color}
\usepackage{mathrsfs}
\usepackage{subfigure}
\usepackage{color}
\usepackage{graphicx}
\usepackage{amsmath,amssymb,amsfonts}
\usepackage{verbatim}

\def\be{\nopagebreak[3]\begin{equation}}
\def\ee{\end{equation}}
\def\ba{\nopagebreak[3]\begin{eqnarray}}
\def\ea{\end{eqnarray}}

\def\lp{\ell_{\rm Pl}}

\def\f{\frac}

\def\rcr{\rho_{\rm max}}
\def\hom{\rm hom}

\def\ul{\underline}

\def\h{\hat}

\def\dd{{\rm d}}
\def\pphi{p_{(\phi)}}

\def\vk{\vec{k}}

\def\Q{\mathcal{Q}}

\def\e{\mathfrak{e}}

\def\q{\mathfrak{q}}

\def\pp{\mathfrak{p}}

\def\T{\mathcal{T}}

\def\b{{\rm b}}
\def\w{\omega}
\def\d{{\rm d}}
\def\q{\mathring{q}}
\def\e{\mathring{e}}

\def\p{\partial}

\def\v{{\rm{V}}}
\def\h{\hat }

\def\T{{\cal T}}

\def\h{\hat }

\def\lp{{\ell}_{\rm Pl}}

\def\q{\mathring{q}}
\def\e{\mathring{e}}
\def\w{\mathring{\omega}}
\def\rmax{\rho_{\mathrm{max}}}

\def\rcr{\rmax}

\usepackage{enumerate}

\def\f{\frac}

\def\dd{\textrm{d}}
\def\d{\textrm{d}}

\def\ul{\underline}

\usepackage{enumerate}

\newcommand{\bmult}{\nopagebreak[3]\begin{multline}}
\newcommand{\emult}{\end{multline}}

\def\h{\hat }

\def\T{{\cal T}}

\def\h{\hat }

\def\b{\mathrm{b}}

\def\lp{{\ell}_{\rm Pl}}

\def\f{\frac}

\def\dd{\textrm{d}}
\def\d{\textrm{d}}

\def\ul{\underline}

\def\rmax{\rho_{\rm max}}

\def\lp{l_{\rm Pl}}

\def\ld{l_{\Delta}}
\def\d{{\rm d}}

\def\ld{\lambda}
\def\v{{\rm v}}
\def\b{{\rm b}}
\def\chilb{\chi(\b,\phi)}
\def\chilx{\chi(x,\phi)}
\def\lam{\lambda}

\def\lu{l_1}
\def\ld{l_2}

\def\lvp{(\lu,\ld,v;\phi)}
\def\vfu{\f{v \pm 4}{v \pm 2}}
\def\vfd{\f{v \pm 2}{v}}
\def\vft{\f{v}{v \pm 2}}

\def\lu{l_1}
\def\ld{l_2}

\def\lvp{(\lu,\ld,v;\phi)}
\def\vfu{\f{v \pm 4}{v \pm 2}}
\def\vfd{\f{v \pm 2}{v}}
\def\vft{\f{v}{v \pm 2}}

\begin{document}
\title{Loop Quantum Cosmology: A brief review\\
}

\author{Ivan Agullo and Parampreet Singh\footnote{agullo@lsu.edu, psingh@lsu.edu 
\\This manuscript is a chapter contribution for a volume edited by A. Ashtekar and J. Pullin, to be published in the World Scientific 
series ``100 Years of General Relativity."  }}%\index[aindx]{Author, F.} % or \aindx{Author, F.}
%\index[aindx]{Author, S.} % or \aindx{Author, S.}

\affiliation{Department of Physics and Astronomy, Louisiana State University, Baton Rouge, LA 70803-4001}

\begin{abstract}
In the last decade, progress on quantization of homogeneous cosmological spacetimes using techniques of loop quantum gravity has led to  
insights on various fundamental questions and has opened new avenues to explore Planck scale physics. These include the problem of 
singularities and their possible generic 
resolution, constructing viable non-singular models of the very early 
universe, and bridging quantum gravity with cosmological observations. This progress, which has resulted from an interplay of sophisticated 
analytical and numerical techniques, has also led to 
valuable hints on loop quantization of black hole and inhomogeneous spacetimes. In this review, we provide a summary of this progress while 
 focusing on concrete examples of the quantization procedure and phenomenology of cosmological perturbations.

%Abstract here
\end{abstract}
%\markright{Customized Running Head for Odd Page} % default is Chapter Title.
%\body

\maketitle

\section{Introduction}\label{sec1}

The goal of this chapter is to apply the techniques of loop quantum gravity (LQG) to cosmological spacetimes. The resulting framework is
known as loop quantum cosmology (LQC). This chapter has a two-fold motivation: to highlight various developments on the theoretical and conceptual issues in the last decade in the framework of loop quantum cosmology, and to demonstrate the way these developments open novel avenues for explorations of Planck scale physics and the resulting phenomenological implications.

From the theoretical viewpoint, cosmological spacetimes provide a very useful stage to make significant progress on many conceptual and technical problems  in quantum gravity. These geometries have the advantage of being highly symmetric, since spatial homogeneity reduces the infinite number of degrees of freedom to a finite number, significantly simplifying the quantization of these spacetimes. Difficult challenges and mathematical complexities still remain, but they are easier to overcome  than in more general situations. The program of canonical quantization of the gravitational degrees of freedom of  cosmological spacetimes dates back to Wheeler and De Witt \cite{wdw1,wdw2}. In recent years, LQC has led to significant insights and progress in quantization of these mini-superspace cosmological models and fundamental questions have been addressed. These include: whether and how the classical singularities are avoided by quantum gravitational effects; how a smooth continuum spacetime emerges from the underlying quantum theory; how do quantum gravitational effects modify the classical dynamical equations; the problem of time and inner product; quantum probabilities; etc.
(see \cite{asrev,lqcreview,agullo-corichi,grainreview2016,ashtekar-barrau,reviewmena} for reviews in the subject). Spacetimes where detailed quantization has been performed include
Friedmann-Lemaitre-Robertson-Walker (FLRW) \cite{aps1,aps2,aps3,apsv,warsaw1,ck2,ck2a,k=-1,szulc,bp,kp1,ap,aps4,rad}, Bianchi \cite{chiou_bianchi,madrid-bianchi,awe2,awe3,we,pswe,ckb9} and Gowdy models \cite{hybrid1,hybrid2,hybrid3,hybrid4}, the latter with an infinite number of degrees of freedom. A coherent picture of singularity resolution and Planck scale physics has emerged based on a rigorous mathematical framework, complemented with powerful numerical techniques. This new paradigm has provided remarkable insights on quantum gravity, and allowed a systematic exploration of the physics of the very early universe. On the other hand, simplifications also entail limitations. Since the formulation and the resulting physics is most rigorously studied in the mini-superspace setting, it is natural to question its robustness when infinite number of degrees of freedom are present, and whether the framework captures the implications from the full quantum theory. The problem of relating a model with more degrees of freedom to its symmetry reduced version is present even at the mini-superspace level. In this setting important insights have been gained on the relation between the loop quantization of Bianchi-I spacetime and spatially flat ($k=0$) isotropic model, which provide useful lessons to relate quantization of spacetimes with different number of degrees of freedom \cite{awe2}. Moreover, the Belinskii-Khalatnikov-Lifshitz (BKL) conjecture \cite{bkl} -- that the structure of the spacetime near the singularities is determined by the time derivatives and spatial derivatives become negligible, which is substantiated by rigorous mathematical and numerical results \cite{berger,dg1}, alleviates some of these concerns and provides a support to the quantum cosmology program.
Finally, recently there has been some concrete progress on the relation between LQC and full LQG, discussed briefly in section 6. From the phenomenological perspective, we are experiencing a fascinating time in cosmology. The observational results of WMAP \cite{wmap7} and  PLANCK \cite{planck}  satellites have provided strong evidence for a primordial origin of the CMB temperatures anisotropies. There is no doubt that the excitement in early universe cosmology is going to continue for several more years, providing a promising opportunity to test implications of quantum gravity in cosmological observations.

This chapter provides a review, including the most recent advances, of loop quantization of cosmological spacetimes and phenomenological consequences. It is organized as follows. Section \ref{sec2} provides a summary of loop quantization of the spatially flat, isotopic and homogeneous model sourced with a massless scalar field. This model was the first example of the rigorous quantization of a cosmological spacetime in LQC \cite{aps1,aps2,aps3}. Because the quantization strategy underlying this model has been implemented for spacetimes with spatial curvature, anisotropies and also in presence of inhomogeneities, we discuss it in more detail. After laying down the classical framework in Ashtekar variables, we discuss the kinematical and dynamical
features of loop quantization and the way classical singularity is resolved and replaced by a bounce. This section also briefly
discusses the effective continuum spacetime description which provides an excellent approximation to the underlying quantum dynamics for
states which are sharply peaked. For a specific choice of lapse, equal to the volume, and for the case of a massless scalar field one
obtains an exactly solvable model of LQC (sLQC) which yields important robustness results on the quantum bounce \cite{acs}. In Sec. 3, we briefly discuss the generalization of loop quantization and the
resulting Planck scale physics to spacetimes with spatial curvature, Bianchi, and Gowdy models. Section \ref{sec4} is devoted to cosmological perturbations. We review the formulation of a quantum gravity extension of the standard theory of gauge invariant cosmological perturbations in LQC. These techniques provide the theoretical arena to study the origin of matter and gravitational perturbations in the early universe. This is the goal of section \ref{sec5} where we summarize the LQC extension of the inflationary scenario and discuss  the quantum gravity corrections to physical observables \cite{aan1,aan2,aan3}. Due to space limitations, it is difficult to cover various topics and details in this chapter. These include the earlier developments in LQC \cite{mb1,mb3,abl}, the path integral formulation of LQC \cite{ach}, entropy bounds \cite{awe1}, consistent quantum probabilities \cite{consistent1,consistent2,consistent3,consistent4,consistent5}, application to black hole interiors
\cite{bh1,bh2,bh3,bh4,bh5,bh6}, and various mathematical \cite{warsaw2,warsaw3,warsaw4,warsaw5} and numerical results
\cite{brizuela,singh-numerical,dgs1} in LQC.  Issues with inverse triad modifications \cite{aps2,aps3}, limitations of the earlier
quantizations in LQC and the role of fiducial scalings \cite{aps3,cs1,cs3}, and issues related to quantization ambiguities and the resulting physical effects \cite{qa1,qa2} are also not discussed. For a review of some of these developments and issues in LQC, we refer the reader to Ref. \cite{asrev} and the above cited references. We  are also unable to cover all the existing ideas to study LQC effects on cosmic perturbations. See \cite{pert_tensor1, ns_inflation, barrau1, barrau2, barrau3, barrau4, bojowald&calcagni, barrau5,
madrid2,madrid3,madrid4,madrid1, wilson-ewing,wilson-ewing2} for different approaches to that problem.  Further information can be found in the chapter ``Loop quantum gravity and observations'' by Barrau and Grain in this volume, and in the review articles
\cite{asrev,lqcreview,calcagni, barraureview, bbcg,reviewmena}. Related to LQC, there have been developments in spin foams and group field
theory, for which we refer the reader to Refs.\cite{vidotto,gftcosmo}.

Our convention for the metric signature is $-+++$, we set $c=1$ but keep $G$ and $\hbar$ explicit in our expressions, to emphasize gravitational and quantum effects. When numerical values are shown, we use Planck units.

\section{Loop quantization of spatially flat isotropic and homogeneous spacetime}\label{sec2}
In this section, we illustrate the key steps in loop quantization of homogeneous cosmological models using the example of spatially
flat FLRW spacetime sourced with a massless scalar field $\phi$. Though simple, this model is rich in physics and provides  a
blueprint for the quantization of models with spatial curvature,  anisotropies and other matter fields. Loop quantization of this
spacetime was first performed in Refs. \cite{aps1,aps2,aps3} where a rigorous understanding of the quantum Hamiltonian constraint, the
physical Hilbert space and the Dirac observables was obtained, and detailed physical predictions were extracted using numerical simulations.
It was soon realized that this model can also be solved exactly \cite{acs}. This feature serves as an important tool to
test the robustness of the physical predictions obtained using numerical simulations. In the following, in Sec. 2.1, we begin with the
quantization of this cosmological model in
the volume representation. We
discuss the classical and the quantum framework, and the main features of the quantum dynamics. We also briefly discuss the effective spacetime description which captures the quantum dynamics in LQC for sharply peaked states to an excellent approximation and provides a very useful arena to understand various phenomenological implications. The exactly solvable model is discussed in Sec. 2.2.

%\subsection{Loop quantization of $k=0$ homogeneous and isotropic model}

\subsection{Loop quantum cosmology: $k=0$ model}
In the following, we outline the classical and the quantum framework of LQC in the spatially flat isotropic and homogeneous spacetime following the analysis of Refs. \cite{aps1,aps2,aps3}. In literature this quantization is also known as `$\bar \mu$ quantization' or `improved dynamics' \cite{aps3}. In the first part we introduce the connection variables, establish their relationship with the metric variables, find the classical Hamiltonian constraint in the metric and the connection variables and obtain the singular classical trajectories in the relational dynamics expressing volume as a function of the internal time $\phi$. This is followed by the quantum kinematics, properties of the quantum Hamiltonian constraint in the geometric (volume) representation, the physical Hilbert space and a summary of the physical predictions. A comparison with the Wheeler-DeWitt theory is also provided both at the kinematical and the dynamical level. An effective description of the quantization performed here, following the analysis of Refs.\cite{jw,vt} is discussed in Sec. 2.1.3.

\subsubsection{Classical framework}
The spatially flat homogeneous and isotropic spacetime is typically considered with a spatial topology $\mathbb{R}^3$ or of a 3-torus
$\mathbb{T}^3$. For the non-compact spatial manifold extra care is needed to introduce the symplectic structure in the canonical framework
because of the
divergence of the spatial integrals. For the non-compact case one introduces a fiducial cell ${\cal V}$, which acts as an infra-red
regulator \cite{aps3}.
Physical implications must be independent of the choice of this regulator, which is the case for the present analysis.\footnote{This is not
true for the earlier quantization in LQC \cite{abl,aps2}, and the lattice refined models \cite{lattice}. For a detailed discussion of these
difficulties in other quantization prescriptions we refer the reader to Refs.\cite{aps3,cs1}.}
Such a cell is not required for the compact topology. %Without a loss of any generality, one can consider any of these topologies. %we
%In the following we consider the compact topology and denote the volume of $\mathbb{T}^3$ with respect to the fiducial metric
%$\q_{ab}$ on the spatial manifold as  $V_o$.
The spacetime metric is given by
\be
\d s^2 = - \d t^2 + a^2 \, \q_{ab} \d x^a \d x^b
\ee
where $t$ is the proper time, $a$ denotes the scale factor of the universe and $\q_{ab}$ denotes the fiducial metric on the spatial manifold.% $\mathbb{T}^3$: $\q_{ab} d x^a d x^b = d x_1^2 + d x_2^2 + d x_3^2$ where $x^a \in [0, \ell_o]$.
With the matter source  as the massless scalar field which serves as a physical clock in our analysis, instead of proper time it is  natural to introduce a harmonic time
$\tau$ satisfying $\Box \tau = 0$ since $\phi$ satisfies the wave equation $\Box \phi = 0$. This corresponds to the  choice of the lapse $N = a^3$. The spacetime metric then becomes
\be
\d s^2 = - a^6 \d \tau^2 + a^2 \, (\d x_1^2 + \d x_2^2 + \d x_3^2) ~.
\ee
In terms of the physical spatial metric $q_{ab} = a^2 \q_{ab}$, the physical volume of the spatial manifold is $V = a^3 V_o$, where $V_o$
is the comoving volume of the fiducial cell in case the topology is $\mathbb{R}^3$, or the comoving volume of $\mathbb{T}^3$ in case the
topology is compact.

Due to the underlying symmetries of this spacetime, the spatial diffeomorphism constraint is satisfied and the only non-trivial constraint is the
 the Hamiltonian constraint. Let us first obtain this constraint in the metric variables. In such a formulation, the canonical pair of
gravitational
 phase space variables consists of the scale factor $a$ and its conjugate $p_{(a)} = - a \dot a$, with `dot' denoting derivative with
respect to the proper time. These variables
  satisfy $\{a, p_{(a)}\} = 4 \pi G/3 V_o$. The matter phase space
 variables are $\phi$ and $p_{(\phi)} = V \dot \phi$, which satisfy $\{\phi, p_{(\phi)}\} = 1$. In terms of the metric variables, the Hamiltonian constraint is given by
 \be \label{hc1} {\cal C}_H =  - \f{3}{8\pi G}\,\f{p_{(a)}^2 V}{a^4} \, + \f{\pphi^2}{2V} \approx 0\, ,
 \ee
 which yields the classical Friedman equation in terms of the energy density, $\rho = p_{(\phi)}^2/2V^2$, for the spatially flat FRW model:
\be\label{classicalfried}
\left(\f{\dot a}{a}\right)^2 = \f{8 \pi G}{3} \rho ~.
\ee
 In order to obtain the classical Hamiltonian constraint in terms of the variables used in LQG: the
 Ashtekar-Barbero SU(2) connection $A^i_a$ and the conjugate triad $E^a_i$, we first notice that due to the symmetries of the isotropic and homogeneous spacetime,
 the connection $A^i_a$ and triad $E^a_i$ can be written as \cite{abl}
 \be
A^i_a \,= \,c \, V_o^{-1/3} \, \w^i_a , ~~~ E^a_i \,= \,p \, V_o^{-2/3} \, \sqrt{\q} \, \e^a_i ~,
\ee
where $c$ and $p$  denote the isotropic connection and triad, and $\e^a_i$ and $\w^i_a$ are the fiducial triads and co-triads compatible with the fiducial metric $\q_{ab}$.
The canonically conjugate pair $(c,p)$ satisfies $\{c,p\} = 8 \pi G \gamma/3$, and is related to the metric variables as $|p| = V_o^{2/3} a^2$ and $c = \gamma V_o^{1/3} \dot a/N$,
where $\gamma$ is the Barbero-Immirzi parameter in LQG, whose value is set to $\gamma \approx 0.2375$ using black hole thermodynamics
\cite{meissner}. The modulus sign over the triad arises because of the two possible orientations, the choice of which does not affect
physics in the absence of fermions.
It is important to note that the above relation between  the triad and the scale factor is true kinematically, whereas the relation between the isotropic connection and the time
derivative of the scale factor is true only for the physical solutions of GR.

It turns out that in the quantum theory, it is more convenient to work with variables $\b$ and $\v$
which are defined in terms of $c$ and $p$ as \cite{acs}:
\be \label{bv} \b := \f{c}{|p|^{\f{1}{2}}}
,\,\,\,\, \v := {\rm sgn}(p) \f{|p|^{\f{3}{2}}}{2\pi G} ,
\ee
where sgn(p) is $\pm 1$ depending on whether the physical and fiducial triads have the same orientation (+), or the opposite (-). The conjugate variables $\b$ and $\v$ satisfy
$ \{\b,\, \v\} = 2\gamma$, and in terms of  which the classical Hamiltonian constraint becomes
\be \label{hc3} {\cal C}_H =  - \f{3}{4\gamma^2}\b^2
|\v|\, + \f{\pphi^2}{4\pi G |\v|} \, \approx 0 \, .\ee
For a given value of $p_{(\phi)}$ and for a given triad orientation,  Hamilton's equations yield an expanding and a contracting
trajectory, given by
\be\label{traj}
\phi = \pm \f{1}{\sqrt{12 \pi G}} \ln \f{\v}{\v_c} + \phi_c
\ee
where $\v_c$ and $\phi_c$ are integration constants. Both trajectories encounter a singularity. In the classical theory, the existence of
a singularity either in past of the expanding branch or in the future of the contracting branch is thus inevitable.

\subsubsection{Quantum framework}
To pass to the quantum theory, the strategy is to promote the classical phase variables and the classical Hamiltonian constraint to their quantum operator analogs.
For the metric variables, this startegy leads to the  Wheeler-DeWitt quantum cosmology. Since, we wish to obtain a loop quantization of the cosmological spacetimes based on
LQG we can not use the same strategy for the connection-triad variables. In LQG, variables used for quantization are the
holonomies of the connection $A_a^i$ along edges, and the
fluxes of the triads along 2-surfaces. (See \emph{Chapter 1}.) For the homogeneous spacetimes, the latter turn out to be proportional to the triad \cite{abl}. The holonomy of the symmetry reduced  connection $A^i_a$ along a straight edge $\e^a_k$ with fiducial length $\mu$ is,
\be
h_k^{(\mu)} \, = \, \cos \left(\f{\mu c}{2}\right) \mathbb{I} + 2  \sin \left(\f{\mu c}{2}\right) \tau_k
\ee
where $\mathbb{I}$ is a unit $2\times2$ matrix and $\tau_k = - i \sigma_k/2$, where $\sigma_k$ are the Pauli spin matrices. Due to the
symmetries of the homogeneous spacetime, the holonomy and flux are thus captured by functions $N_\mu (c) := e^{i \mu c/2}$ of $c$, and the triads $p$ respectively. Since $\mu$ can take arbitrary values, $N_\mu$ are \emph{almost periodic functions} of the connection $c$. The next task is to find the appropriate representation of the abstract $\star$-algebra generated by almost periodic functions
\footnote{A continuous function $F$ of an unrestricted real variable $x$ is almost periodic if $F(x + \tau) = F(x)$
holds to an arbitrary accuracy for infinitely many values of $\tau$, such that translations $\tau$ are spread over the whole real line
without arbitrarily large intervals \cite{bohr}.}
of the connection $c$: $e^{i \mu c/2}$, and the triads $p$. It turns out that there exists a unique kinematical representation of
algebra generated by these functions in LQC \cite{unique1,unique2,unique3}. This result has parallels with
existence of a unique irreducible representation of the holonomy-flux algebra in full LQG \cite{lost,cf}.  The gravitational sector of the kinematical Hilbert space ${\cal H}_{\rm{kin}}$ underlying this representation in LQC is a space of square integrable functions on the Bohr compactification of the real line: $L^2(\mathbb{R}_{\rm{Bohr}},\d \mu_{\rm{Bohr}})$ \cite{abl}. Use of holonomies in place of 
connections does not directly affect the matter sector. For this reason, the matter sector of the kinematical Hilbert space is obtained by
following the methods in the Fock quantization.%
\footnote{Polymer quantization of matter sector in a similar setting has been studied in some of the works, see for eg. \cite{polymer1,polymer2,polymer3}.}

It is important to note the difference between the gravitational part of ${\cal H}_{\rm{kin}}$, and the one obtained by following the Wheeler-DeWitt procedure where the gravitational part of the kinematical Hilbert space is $L^2(\mathbb{R},\d c)$. In LQC, the normalizable states are the countable sum of $N_\mu$, which satisfy: $\langle N_\mu|N_\mu' \rangle = \delta_{\mu \mu'}$, where $\delta_{\mu \mu'}$ is a Kronecker delta. This is in contrast to the Wheeler-DeWitt theory where one obtains a Dirac delta. Thus, the kinematical Hilbert space in LQC is fundamentally different from one in the Wheeler-DeWitt theory. The  intersection between the kinematical Hilbert space in LQC and the Wheeler-DeWitt theory consists only of the zero function. Since the system has only a finite degrees of freedom, one may wonder why the the von-Neumann uniqueness theorem, which leads to a unique Schr\"odinger representation
in quantum mechanics, does not hold. It turns out that for the theorem to be applicable in LQC,  $N_\mu$ should be weakly continuous in $\mu$. This condition is not met in LQC, and the von-Neumann theorem is bypassed. (For further details on this issue, we refer the reader to Ref. \cite{shadow}).

The action of the operators $\hat N_\mu$ and $\hat p$ on states $\Psi(c)$ is by multiplication and differentiation respectively. On the
states in the triad representation labelled by eigenvalues $\mu$ of $\hat p$ , the
action of $\hat N_\mu$ is translational:
\be
\hat N_\zeta \, \Psi(\mu) = \Psi(\mu + \zeta),
\ee
where $\zeta$ is a constant%
\footnote{Note that we have used $N_\zeta$ instead of $N_\mu$ to avoid confusion with the argument of the
wavefunction $\Psi(\mu)$.}, 
and $\hat p$ acts as:
\be
\hat p \, \Psi(\mu) = \f{8\pi \gamma \lp^2}{6}\, \mu \Psi(\mu) ~.
\ee
Before we proceed to the quantum Hamiltonian constraint, we note that the change in the orientation of the triads which does not lead to any physical consequences in the absence of fermions corresponds to a large gauge transformation by a parity operator $\hat \Pi$ which acts on $\Psi(\mu)$ as: $\hat \Pi \Psi(\mu) = \Psi(-\mu)$. The physical states
in the absence of fermions are therefore required to be symmetric, satisfying $\Psi(\mu) = \Psi(-\mu)$.

To obtain the dynamics in the quantum theory, we start with the Hamitonian constraint in full LQG in terms of triads $E^a_i$ and the field strength of the connection $F_{ab}{}^{k}$:
\footnote{The Hamiltonian constraint consists of two terms proportional to $\epsilon_{i j k} F^i_{ab} E^{a j} E^{bk}$ and $K^i_{[a}K_{b]}^jE^a_iE^b_j$, where $K^i_a$ capture the extrinsic curvature. These two terms turn out to be proportional to each other for the spatially flat homogeneous and isotropic model. Eq.(\ref{cgrav}) captures the resulting total contribution.} 
\be \label{cgrav} C_{\mathrm{grav}} = - \gamma^{-2}\,
\textstyle{\int}_{\cal C}\,\, \dd^3 x\,\big[N ({\det
q})^{-\f{1}{2}}\,\epsilon^{ij}{}_k E^a_i
E^b_j\big]\,\, F_{ab}{}^k
\ee
which in terms of the symmetry reduced triads and lapse $N=a^3$ becomes,
\be
C_{\rm{grav}} \, = \, \gamma^{-2} \, V_o^{-1/3} \, \epsilon^{i}_{~j k} \, \e^a_i \, \e^b_j \, |p|^2 F_{ab}{}^{k} ~.
\ee
The field strength $F_{ab}^{~k}$ is expressed in terms of the holonomies over a square plaquette $\Box_{ij}$ with length $\bar \mu V_o^{1/3}$ in the $i-j$ plane spanned by fiducial triads:
\be
F_{ab}^{~k} = - 2 \, \lim_{Ar \Box \rightarrow 0} {\rm{Tr}} \left(\f{h_{{\Box}_{ij}} - \mathbb{I}}{Ar\Box} \, \tau^k\right) \, \w^i_a \, \w^j_b ~.
\ee
Here Ar$\Box$ denotes the area of the square plaquette, and $
h_{\Box_{ij}} = h_i^{(\bar \mu)} h_j^{(\bar \mu)} (h_i^{\bar \mu})^{-1} (h_j^{\bar \mu})^{-1}$,
with $\bar \mu$ denoting the edge length of the plaquette. Note that due to the underlying quantum geometry, the  limit Ar$\Box \rightarrow 0$ does not exist. Instead one has to shrink the area of the loop to the minimum non-zero eigenvalue of the area operator in LQG. We denote this minimum area to be $\Delta \lp^2$ where $\Delta = 4 \sqrt{3} \pi \gamma$\cite{awe2}. This results in the following
functional dependence of $\bar \mu$ on the triad \cite{aps3}
\be\label{mubar}
\bar \mu^2 = \f{\Delta \lp^2}{|p|} ~,
\ee
where we have used the expression for the physical area of the loop which equals $\bar \mu^2 |p|$.
Due to this form of $\bar \mu$, the action of $N_{\bar \mu}$ on the triad eigenstates is not by a simple translation. However, switching to the volume representation gives the
simple translation action, and therefore in the quantum theory it is more convenient to work with this representation in which the action of the conjugate operator $\widehat{\exp(i \lambda \b)}$ (with $\lambda^2 = \Delta \lp^2$) and the volume operator is:
\be
\widehat{\exp(i \lambda \b)} \, |\nu\rangle \, = \, |\nu - 2 \lambda \rangle, ~~~~ \hat V \, |\nu\rangle \, = \, 2 \pi \gamma \lp^2 \, |\nu| \, |\nu\rangle ~~~  ~
 \ee
 where $\nu = \v/\gamma \hbar$. Using these operators, we can find the solutions to $\hat C_H \Psi(\nu,\phi) = \hat C_{\rm{grav}} + 16 \pi G \hat C_{\rm{matt}} \Psi(\nu,\phi) = 0$. For the massless scalar field as the matter source, the quantum constraint equation results in the following:
\be \label{cons1}
\partial_\phi^2 \, \Psi(\nu,\phi) \, = \, 3 \pi G \, \nu \, \f{\sin \lambda \b}{\lambda} \nu \, \f{\sin \lambda \b}{\lambda} \, \Psi(\nu,\phi) \, =: \, - \, \Theta \Psi(\nu,\phi) \\
\ee
where $\Theta$ is a positive definite, second order difference operator:
\be\label{thetacons}
\Theta \Psi(\nu, \phi) \, := \, - \f{3 \pi G}{4 \lambda^2} \, \nu \left((\nu + 2 \lambda) \Psi(\nu + 4 \lambda) - 2 \nu \Psi(\nu,\phi) + (\nu - 2 \lambda) \Psi(\nu - 4 \lambda)\right) ~.
\ee
The form of the quantum constraint turns out to be very similar to the Klein-Gordon theory, where $\phi$ plays the role of time and $\Theta$ acts like a spatial Laplacian operator.
As in the Klein-Gordon theory, the physical states can be either positive or the negative frequency solutions. Without any loss of generality we choose  the physical states to
be solutions of the positive frequency square root of the quantum constraint:
\be \label{qhc5} -i\, \p_\phi \Psi(\nu,\phi) = \sqrt\Theta\,
\Psi(\nu,\phi) ~. \ee
The inner product for these physical states can be obtained using group averaging \cite{dm,almmt,abc}, and is given by
\be
\langle \Psi_1| \Psi_2\rangle \, = \, \sum_\nu \, \bar \Psi_1(\nu,\phi_o) |\nu|^{-1} \Psi(\nu_2,\phi_o) ~.
\ee

To extract physical predictions, we  introduce Dirac observables which are self-adjoint with respect to the above inner product. One of the Dirac observables is
$\hat p_{(\phi)}$ which is a constant of motion. The other is $\hat V|_\phi$, the volume at internal time $\phi$. On states $\Psi(\nu,\phi)$, the action of these observables is
\be
\hat V|_{\phi_{o}} \Psi(\nu,\phi) \, =  \, 2 \pi \gamma \lp^2 \, e^{i \sqrt{\Theta} (\phi - \phi_o)} |\nu| \Psi(\nu,\phi_o)
\ee
and
\be \hat p_\phi \Psi(\nu,\phi) \, = \, - i \hbar \, {\p_\phi} \, \Psi(\nu,\phi) = \hbar \sqrt{\Theta} \Psi(\nu,\phi) ~.
\ee
%Similarly, one can define a Dirac observable corresponding to the energy density of the scalar field.
Note that the Dirac observables preserve the positive and negative frequency subspaces. The symmetric wavefunctions which satisfy
eq.(\ref{thetacons}) have support on a lattice $\nu = \pm \epsilon + 4 n \lambda$ with $\epsilon \in [0,4 \lambda)$. Any subspace spanned by
the wavefunctions labelled by $\epsilon$ is preserved under evolution and the action of the Dirac observables. Therefore, there is a
superselection and it suffices to consider states with a particular value of $\epsilon$. Further, physical predictions are insensitive to
the choice of the lattice parameter. In the following analysis we choose $\epsilon = 0$, since this choice of lattice parameter results in
the possibility of the evolution encountering the classical singularity at the zero volume. Any other value of $\epsilon$ can also be
chosen, say $\epsilon = 0.1$, however in such case zero volume does not lie on support of the eigenfunctions of the $\Theta$ operator.

Before we discuss some of the key features of the quantum Hamiltonian constraint in LQC and the resulting physics, we note that a similar analysis goes through for the
Wheeler-DeWitt theory based on the metric variables. At a
 kinematical level, the Wheeler-DeWitt Hilbert space consists of wavefunctions $\ul{\Psi}(a,\phi)$ on which the scale factor and $\phi$ operators act multiplicatively, and
 the operators corresponding to their conjugate variables act as differential operators. The physical states are found by promoting ${\cal C}_H$ (\ref{hc1}) to an operator and solving
 $\hat {\cal C}_H \Psi(a,\phi) = 0$. The resulting quantum constraint turns out to be a differential equation \cite{aps3,acs}:
 \be\label{wdwcons}
 \p_\phi^2 \ul{\Psi}(z,\phi) = 12\pi G\, \partial_z^2 \ul{\Psi}(z,\phi) =: - \ul\Theta  \, \ul\Psi(z,\phi)\,
 \ee
 where $z = \ln a^3$ and $\ul\Theta$ is the evolution operator in Wheeler-DeWitt theory.  This brings out another
fundamental difference between the Wheeler-DeWitt theory and LQC. Unlike the Wheeler-DeWitt theory, the quantum constraint in LQC is a
discrete operator with discreteness determined by the underlying quantum geometry in LQG. For the scales where the spacetime curvature is
very small compared to the Planck scale, which corresponds to  the large volumes for the present model, the $\Theta$ operator in LQC
approximates the $\ul\Theta$ operator in the Wheeler-DeWitt theory \cite{abl,aps2}. Thus, the continuum differential geometry is recovered
from the underlying discrete quantum geometry at the small spacetime curvature.

 \begin{figure}[]
  \begin{center}
    \includegraphics[width=3.2in,angle=0]{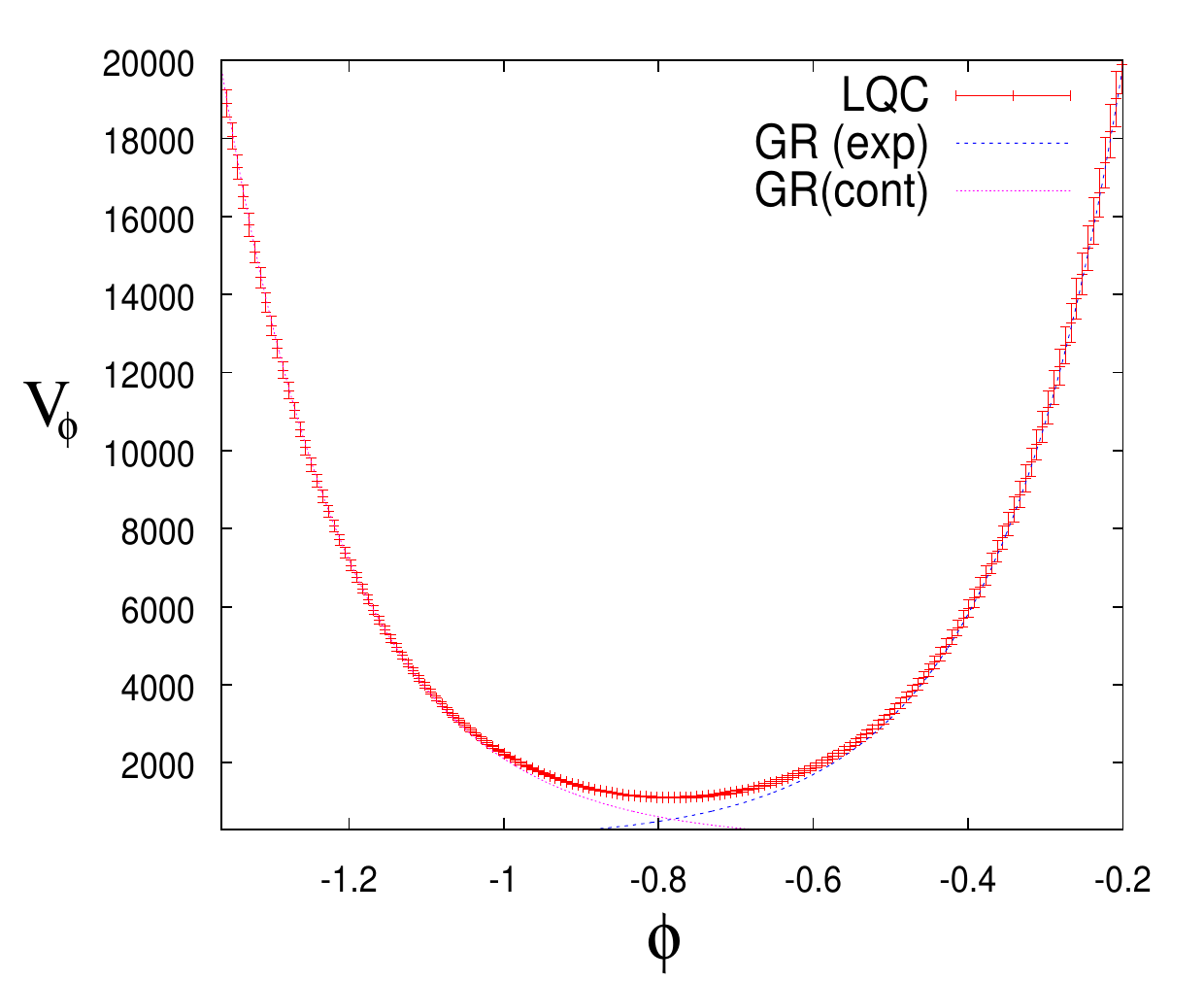}
 \caption{A comparison of the quantum evolution in LQC for the volume observable (along with its dispersion) and the classical trajectories is shown. Unlike the general relativistic trajectories which lead to a singularity in the future evolution for the contracting branch and the past evolution for the expanding branch, the LQC trajectory is non-singular. The LQC trajectory bounces in the Planck regime and the loop quantum universe evolves in a non-singular way. The dispersions across the bounce are correlated, and their asymmetry depends on the method of initial state construction (see Ref. \cite{aps2} for details of different methods). The state retains its peakedness properties in the above evolution, since the relative dispersion approaches a constant value at large volumes.}
 \end{center}
 \end{figure}

 Quantum evolution of physical states can be studied numerically using the quantum constraint eq.(\ref{thetacons}). One considers an initial state far away from the Planck regime, with large volumes peaked at a certain value of $p_{(\phi)}$ at a classical trajectory. Recall that in the classical theory, for a given value of $p_{(\phi)}$ there exists an expanding and a contracting trajectory which are disjoint and singular. In numerical simulations, the state can be either chosen such that it is peaked on the expanding trajectory at late times or on the contracting trajectory at early times. Using $\phi$ as a clock, such a state, say chosen peaked on the expanding trajectory, is then numerically evolved towards the classical  big bang singularity. The first numerical simulations were carried out using sharply peaked Gaussian states \cite{aps2,aps3}. Such states were shown to
 remain sharply peaked on the classical expanding trajectory for a long time in the backward evolution, till the spacetime curvature reaches approximately a percent of the Planck curvature. At the higher curvature scales, departures between the classical trajectory and quantum evolution become significant, and the loop quantum universe bounces when the energy density reaches a
  maximum value $\rho_{\rm{max}} \approx 0.41 \rho_{\rm{Pl}}$ \cite{aps3}.   After the bounce, the quantum evolution is such that the state becomes sharply peaked on the classical contracting trajectory. Quantum gravitational effects thus bridge the two singular  classical trajectories providing a non-singular evolution avoiding the classical singularity.  A result of the variation of volume with respect to internal time from a  typical simulation is illustrated in Fig. 1 where the LQC evolution is also compared with the two classical trajectories in general relativity (GR). It is clearly seen that the quantum geometric effects play role only near the bounce and quickly become negligible when spacetime curvature becomes small. These studies have been recently generalized for very widely spread states and highly squeezed and non-Gaussian states which capture the evolution of more quantum universes \cite{dgs2,squeezed}, using high performance computing and faster algorithms \cite{dgs1}. The results of quantum
bounce are found to be
robust for all types of states. The existence of bounce does not require any fine tuning of the parameters or any special conditions. The quantum bounce is also found to be robust for slightly different quantization prescriptions in LQC \cite{mmo,mop}.
 In contrast to the loop quantum evolution, the quantum evolution of Wheeler-DeWitt states yields a strinkingly different picture. Initial states peaked on a classical trajectory remained  peaked throughout the evolution and encounter the classical singularity. For the Wheeler-DeWitt states, the expectation values of the volume observable lie on the classical trajectory for all values of $\phi$.

 Thus, we find that unlike in the Wheeler-DeWitt theory, in LQC classical singularities are replaced by the bounce. The existence of bounce is tied to the underlying discrete quantum
 geometry -- a feature which is absent in the Wheeler-DeWitt theory. The quantum evolution for various states in LQC illustrates the way classical GR is recovered in the low curvature regime. Thus, LQC not only provides a non-singular ultra-violet extension of the classical cosmological models, but also leads to the desired infra-red limit.
Finally, we note that this feature provides an important  criterion to single out the $\bar \mu$ quantization as performed in the above analysis of the various possible
choices \cite{cs1,cs3}. In particular, it is useful to note that in the earlier quantization of LQC, called the $\mu_o$ scheme in literature, edge lengths of the loop over which holonomies were constructed were considered to be constant \cite{abl,aps2}. It does not yield the correct infra-red limit and can lead to `quantum gravitational effects' at arbitrarily small spacetime curvatures \cite{aps2,aps3,cs1}. These difficulties are shared by the lattice refined models \cite{lattice}. It is interesting to note that the conclusion that $\bar \mu$ quantization \cite{aps3} is the only  consistent quantization in LQC is not solely tied to the infra-red limit of the theory. This conclusion can also be reached by demanding that the physical predictions be invariant under the rescalings of the fiducial cell \cite{cs1}, by demanding the stability of the quantum difference equation \cite{khanna,brizuela,tanaka,singh-numerical}
and by demanding that the factor ordering ambiguities in gravitational part of the quantum constraint in LQC disappear in the limit where Wheeler-DeWitt theory is approached \cite{nelson-marie}. All these independent arguments provide a robust understanding of the viability of the $\bar \mu$ quantization of the cosmological spacetimes.

\subsubsection{Effective spacetime description\label{sec.eff}}

In Sec. 2.1.2 we discussed the way underlying quantum geometry in LQG results in a quantum difference equation in LQC. Evolution of states
with this difference equation predict a quantum bounce at the Planck scale. The question we are interested in now is whether it is possible to capture the key features of the quantum evolution, including the quantum bounce, in a continuum spacetime description at an effective level, allowing an $\hbar$-dependence in the metric coefficients. If so, is it possible to obtain a modified differential Friedman and Raychaudhuri equations which incorporate the leading quantum gravitational effects? If such a set of reliable quantum-gravity-corrected equations exist, then the exploration of phenomenological implications would become much simpler numerically, within the approximations and caveats underlying these equations. Note however that, while physics obtained from such an effective spacetime description can provide important insights on the underlying quantum geometry, it is imperative to rigorously confirm the implications using full quantum dynamics in LQC where ever possible. It turns out that for states which satisfy certain semi-classicality requirements and lead to a universe which is macroscopic at late times, an effective continuum spacetime description of the loop quantum dynamics can indeed be derived using a geometrical formulation of quantum mechanics \cite{aats,aschilling}. It provides  an effective Hamiltonian from which a modified Friedman equation can be obtained. The result is an effective dynamical trajectory which turns out to be in an excellent agreement with the quantum dynamics for the sharply peaked states \cite{aps2,aps3,dgs2}. In the following we briefly summarize the underlying method of deriving the effective dynamics following the analysis of Refs. \cite{jw,vt}, obtain the modified Friedmann equation for the massless scalar field model, and discuss its main features. Various phenomenological implications of this modified Friedmann dynamics have been extensively discussed in the literature (see Ref. \cite{asrev} for a review), a couple of which will be discussed briefly in Sec. 3.

In the geometrical formulation of quantum mechanics \cite{aats,aschilling}, one treats the space of quantum states as an infinite dimensional phase space $\Gamma_Q$. The symplectic form ($\Omega_Q$) on the phase space is given by the imaginary part of the Hermitian inner product on the Hilbert space. The real part of the inner product determines a Riemannian metric on $\Gamma_Q$. One then seeks a  
relation between the quantum phase space $\Gamma_Q$ with symplectic structure $\Omega_Q$ and the classical phase space $\Gamma$ with 
symplectic structure $\Omega$. The relation is given by an embedding of the finite dimensional classical phase space onto $\bar \Gamma_Q \subset \Gamma_Q$. To capture the Hamiltonian flow on $\Gamma_Q$ that generates full quantum dynamics, one must find an {\it astute} embedding such that the quantum  Hamiltonian flow is tangential to $\bar \Gamma_Q$ to a high degree of approximation. This requirement is very non-trivial and there is no guarantee that such an embedding can be found. However, if such an embedding exists then the projection of the quantum Hamiltonian flow on $\bar\Gamma_Q$ provides the quantum corrected trajectories that capture the main quantum effects to a high degree of approximation. $\bar \Gamma_Q$ is by construction isomorphic with the classical phase space $\Gamma$ and a point  in $\Gamma$ is labelled by $\xi^{o} \equiv (q_i^o,p_i^o)$. Therefore the quantum state corresponding to a point of $\bar\Gamma_Q$ is denoted by $\Psi_{\xi^o}$. A required embedding must satisfy $q_i = \langle \Psi_{\xi^o},\, \hat q_i \Psi_{\xi^o} \rangle$, and $p_i = \langle \Psi_{\xi^o},\, \hat p_i \Psi_{\xi^o} \rangle$.  To find a suitable embedding, one makes careful choice of appropriate states $\Psi_{\xi^o}$, such as coherent states, by choosing appropriate parameters, such as fluctuations. Once $\bar \Gamma_Q$ is found, the leading quantum corrections are well-captured in terms of the classical phase space variables. By carrying out this procedure, one thus obtains modified classical dynamical equations (the effective equations), which incorporate quantum gravitational corrections via a controlled approximation in terms of the parameters of the state. This approach to effective equations is called the `embedding method'

Another approach is the `truncation method' where one introduces a coordinate system on $\bar \Gamma_Q$ using the expectation values $(\bar q_i, \bar p_i)$, fluctuations and the higher order moments \cite{mb_as_eff1,mb_as_eff2}. The quantum Hamiltonian flow on the Hilbert space yields a set of coupled non linear differential equations which are infinite in number for all the moments. By suitably truncating this set up to a finite number of terms, one can then obtain classical dynamical equations with quantum corrections up to the truncated order. In comparison to the embedding approach where appropriate states and their parameters need to be chosen carefully to obtain approximately tangential Hamiltonian vector field, the truncation method is more systematic. However, it is difficult to understand the role played by the infinite number of moments which are truncated out, and the error associated with this truncation.

In LQC, effective equations have been derived using the embedding \cite{jw,vt,psvt} as well the truncation method \cite{mb_as_eff1,mb_as_eff2}. However, since most of the numerical studies on confirming the validity of the effective dynamics and 
phenomenological implications have been performed for the embedding method, we will focus on this approach. For the massless scalar 
field, the effective Hamiltonian constraint, up to the approximation where terms proportional to the square of the quantum fluctuations of 
the state,  using the embedding method is found to be \cite{vt}:
 \be\label{effham1} {C}^{\mathrm{(eff)}}_H \, = \, -\f{3
\hbar}{4 \gamma \lam^2} \nu \, \sin^2(\lam \b) \, + \, \f{1}{4 \pi \gamma \lp^2}\f{p_{(\phi)}^2}{\nu} ~.
\ee
Physical solutions satisfy ${C}^{\mathrm{(eff)}}_H \approx 0$, which yields
\be\label{effham2}
\f{3}{8 \pi G \gamma^2 \lambda^2} V \sin^2(\lam \b) = \f{p_{(\phi)}^2}{2 V} ~.
\ee
The modified Friedmann and Raychaudhuri equations can be found using Hamilton's equation for $V$ and $\b$ respectively, which satisfy $\{\b, V\} = 4 \pi G \gamma$. As an example,
 Hamilton's equation for volume gives,
\be
\dot V = \{V,{C}^{\mathrm{(eff)}}_H\} = - 4 \pi G \gamma \f{\p}{\p \b} {C}^{\mathrm{(eff)}}_H = \f{3}{\gamma \lambda} \sin(\lambda \b) \cos(\lambda \b) ~ ,
\ee
from which it is straightforward to derive the modified Friedmann equation for the Hubble rate $H = \dot V/3 V$ using eq.(\ref{effham2}):
\be \label{fried} H^2 = \f{8 \pi G}{3} \, \rho \left(1 -
\f{\rho}{\rcr}\right) ~~~ \mathrm{with} ~~~ \rcr = \f{3}{8 \pi G \gamma^2 \lam^2} ~. \ee
The quantum gravitational correction thus appears as a $\rho^2$ modification to the classical Friedmann equation (\ref{classicalfried}),  
with a negative sign.\footnote{The $\rho^2$ modification
albeit with a positive sign in front of $\rho/\rho_{\rm max}$ also appears in brane world scenarios in string cosmological models. For the 
modification to be negative one requires one of the extra dimensions to be time-like \cite{sahni}. For a comparative analysis of the 
properties of the above modified Friedmann equation in LQC with the braneworld scenarios see Ref. \cite{ps06,ps-soni1}.} The modified 
Raychaudhuri equation%
\footnote{An interesting relation between this modified Raychaudhuri equation and the resulting structure of the canonical phase space has been explored as an inverse problem \cite{ps-soni1}, without any a priori assumptions about the Hamiltonian framework. It has been suggested that existence of Raychaudhuri equation with such modifications, quadratic or higher order in energy density, requires holonomies of the connection as phase space variables.} 
can be similarly derived from the Hamilton's equation for $\dot b$, which yields
\be\label{raych} \f{\ddot a}{a} = - \f{4 \pi G}{3} \, \rho \,
\left(1 - 4 \f{\rho}{\rcr} \right) - 4 \pi G \, P \, \left(1 -
2 \f{\rho}{\rcr} \right) . ~ \ee
where $P$ denotes the pressure which is equal to the energy density for the massless scalar field model. The energy-matter 
conservation law remains unchanged from the classical theory.
\footnote{Strictly speaking this is true only when the Hamiltonian constraint does not contain any terms which include the inverse triad modifications which due to the choice of lapse are absent in our case. For models where such modifications can be consistently incorporated, such as in the $k=1$ model, the conservation law is also modified 
\cite{ps05,joao}.}

From eq. (\ref{fried})  one concludes that the scale factor of the universe bounces when $\rho = \rcr \approx 0.41 
\rho_{\rm{Pl}}$. Unlike the classical theory, the Hubble rate does not grow unboundedly through out the evolution, but is bounded above by 
a maximum value $|H|_{\rm{max}} = 1/(2 \gamma \lambda)$ which occurs at $\rho = \rcr/2$. Note that the effective dynamics predicts the bounce at the same value of energy density which is found to be the supremum of the expectation values of the energy density observable $(\rho_{\rm{sup}})$ in exactly solvable LQC (as we shall see in eq.(\ref{rhosup})), and the value observed in various numerical simulations for the sharply peaked Gaussian states \cite{aps3,dgs2}. For such states, the effective dynamical trajectory is in excellent agreement with the quantum evolution at all scales. This may seem surprising
because the initial semi-classical state used to derive the effective dynamics is chosen in the regime where quantum gravitational effects 
are negligible and near the bounce some of the underlying assumptions on the parameters of the state can be suspect \cite{vt}.\footnote{It 
has been argued that for the non-compact topology, in the limit of removal of infra-red regulator quantum fluctuations do not affect the 
effective Hamiltonian \cite{rovelli}. However, this argument does not provide an answer to the puzzle about the validity of the effective 
dynamics at the bounce as discussed in the literature \cite{vt}, since the apparently failing assumptions which are problematic are 
fluctuation independent \cite{psvt}.} A careful analysis of the underlying assumptions in the derivation of effective dynamics shows that they are satisfied all the way up to the bounce for the sharply peaked states \cite{psvt}. Thus, for such states effective dynamics provides a very reliable continuum spacetime description of this model. 
For states which have large relative fluctuations or have very large non-Gaussianity, numerical simulations find departures between the 
quantum evolution and the effective dynamics obtained from eqs.(\ref{fried}) and (\ref{raych}) \cite{dgs2,squeezed}. Interestingly, it turns out that the above effective dynamics always overestimates the energy density at the bounce. This observation is consistent with the result in sLQC that the maximum value of the expectation value of energy density $\rho_{\rm{sup}}$ is  same as $\rho_{\rm{max}}$,  and the bounce density for certain states in the quantum theory can be smaller \cite{cm1,cm2}. An insight from sLQC is that for the case of a spatially flat model with a massless scalar field, the modified Friedman equation (\ref{fried}) can be generalized to arbitrary states in the physical Hilbert space with $\rcr$ replaced with the expectation value of the energy density obseravble at the bounce \cite{aabg}.

Effective dynamics has provided many important insights on the physics at the Planck scale in LQC. Using effective equations, a 
relationship of effective Hamiltonian in LQC with a covariant effective action containing infinite number of higher order curvature terms 
has been explored \cite{palatini}. An extensive understanding is reached on genericity of singularity resolution and occurence of 
inflation. For generic matter the above bound on the Hubble rate
leads to the resolution of strong curvature singularities \cite{ps09,ps11,ps15,ss1}, and the bounce dynamics plays an important 
role to make the probability for inflation close to unity in LQC \cite{as,as3,ck-inflation,corichi-sloan}. We discuss some of the 
aplications of effective dynamics in Sec. 3. For a more complete discussion of various phenomenological implications we refer the reader to Ref. \cite{asrev}.

 \subsection{Solvable Loop Quantum Cosmology (sLQC)}
The spatially flat loop quantum cosmological model with a massless scalar field can be solved exactly by passing to
the $\b$ (the conjugate to volume) representation \cite{acs}. The exact solvability of this model proves extremely important to test the robustness of various physical implications obtained in Sec. 2.1. Another advantage of this analysis is that similarities and differences between LQC and the Wheeler-DeWitt theory become very transparent. In both frameworks, the underlying exactly soluble models are very similar, such as in the form of the quantum constraint and the action of momentum observable. But, there are also some important distinctions, in particular on the behavior of the expectation values of the volume. This difference is pivotal in proving some important results, including the genericness of quantum bounce in sLQC and the occurence of the singularity in the Wheeler-DeWitt theory. Due to its simplicity and powerful features, sLQC has been widely applied in different settings to gain important insights on different problems in LQC, e.g., \, (i) to understand the growth of fluctuations for various states across the bounce \cite{cs2,kp2,cm1,cm2},\, (ii) to develop a path integral formulation of LQC to understand various conceptual issues and explore links with spin foam models \cite{ach,ach2,ach3},\, and,\, (iii) to understand the quantum probabilities for the occurence of the bounce in LQC \cite{consistent2,consistent3} and singularities in Wheeler-DeWitt theory \cite{consistent1} using the consistent histories framework \cite{griffiths,hartle}. Due to space limitations it is not possible for us to elaborate on these various interesting applications in detail, and refer the interested reader to the review \cite{asrev}.

In LQC, since the wavefunctions in the volume representation have  support on a discrete
interval $\nu = 4 n \lambda$, the wavefunctions $\Psi(\b,\phi)$ in the $\b$ representation  have support on
the continuous interval $(0, \pi/\lambda)$. In contrast, in the Wheeler-DeWitt theory $\b\in (-\infty, \infty)$. In both the theories,
 the quantum Hamiltonian constraint in the $\b$ representation is a differential equation. Let us start with sLQC, where it is given by
 \be
\p_\phi^2 \, \chilb = 12 \pi G \left(\f{\sin \lambda \b}{\lambda} \p_{\b}
\right)^2 \, \chilb \, .\ee
It is convenient to change the variable to $x$ with $x \in (-\infty,
\infty)$:
%As in the case of the \WDW theory, we can make a change of
%coordinates to rewrite this constraint as a Klein-Gordon equation:
%in terms of the variable $x$,
%
\be x = \f{1}{\sqrt{12 \pi G}} \, \ln\left(\tan \f{\lambda \b}{2}\right) ~,
\ee
using which the quantum Hamiltonian constraint takes a very simple form of the wave equation,%
\be\label{lqc_bcons} \p_\phi^2 \, \chi(x,\phi)  = \p_x^2\, \chilx =: - {\Theta} \,
\chilx .\ee
As in the case of the volume representation, the physical Hilbert space consists of the
positive frequency solutions which satisfy $
-i\p_\phi\chi(x,\phi) = \sqrt{\Theta}\chi(x,\phi)$.
Further, the requirement that the physics be invariant under the change of the
orientation of the triads leads to $\chi(x,\phi) =
-\chi(-x,\phi)$. Due to this antisymmetric condition, it turns out that
every solution $\chilx$ can expressed in terms of the right $(x_-)$ and the left moving $(x_+)$
parts $\chilx = \f{1}{\sqrt{2}} (F(x_+) - F(x_-))$
where $x_\pm = \phi\pm x$. The physical inner product can be obtained in terms of the left moving (or the right moving) part as:
\be
(\chi_1,\chi_2)_{\mathrm{phys}} = - 2 i \int_{-\infty}^\infty \, \dd
x \bar F_1(x_+) \p_x F_2(x_+) ~. \ee

A similar construction can be carried out for the Wheeler-DeWitt theory, where, unlike sLQC, the resulting wavefunctions
$\chi(y,\phi)$, where $y = (12 \pi G)^{-1/2} \ln(\b/\b_o)$ with $\b_o$ an arbitary constant, are not subject to the requirement that they must have support on the
left and right moving sectors. In the Wheeler-DeWitt theory, these sectors decouple and one can 
choose wavefunctions composed solely of the left moving or the right moving solutions.
Otherwise, the form of the quantum constraint, and the action of the momentum observable are 
identical. However, a crucial difference appears in the expectation value of the volume observable. For sLQC it turns out to be
\be\label{slqc_vol} (\chi, \, \hat V|_\phi
\chi)_{\mathrm{phy}} \, = \, 2 \pi \gamma \lp^2 \, (\chi,
|\hat\nu|_\phi \chi)_{\mathrm{phy}} \, = \, V_+ e^{\sqrt{12 \pi G}
\phi} + V_- e^{-\sqrt{12 \pi G} \phi} . \ee
Here $V_\pm$ are positive constants determined by the initial state:
\be V_{\pm} = \f{4 \pi
\gamma \lp^2 \lambda}{\sqrt{12 \pi G}} \int_{-\infty}^\infty \d x_+
\left|\f{\d F}{\d x_+}\right|^2 e^{\mp \sqrt{12 \pi G} x_+} . \ee
In contrast, for the Wheeler-DeWitt theory the expectation value of $\hat V|_\phi$ for the left moving sector
are
\be\label{wdw_vol1} (\ul \chi_L,\, \hat V|_\phi \ul
\chi_L)_{\mathrm{phy}} = 2 \pi \gamma \lp^2 \, (\ul \chi_L,
|\h\nu|_\phi \ul \chi_L)_{\mathrm{phy}} = V_* e^{\sqrt{12 \pi G} \phi}
\ee
with
\be V_* = \f{8 \pi \gamma \lp^2}{\sqrt{12 \pi G} \b_o}
\int_{-\infty}^\infty \dd y_+ \left|\f{\dd \ul \chi_L}{\dd
y_+}\right|^2 \, e^{-\sqrt{12 \pi G} y_+} ~. \ee

In the Wheeler-DeWitt theory, the expectation values $\langle \hat V|_\phi\rangle$ approach zero as $\phi \rightarrow -\infty$. The left moving modes, which
correspond to the expanding trajectory, thus encounter a big bang singularity in the past. Similarly, the right moving modes encounter a big crunch singularity
in the future evolution. Note that this conclusion does not assume any profile of the initial state in this theory. 
An analysis of the quantum probabilities using
consistent histories approach shows that the probability for a singularity to occur in this Wheeler-DeWitt model in asymptotic past of future is unity \cite{consistent1}, even for the states composed of the arbitrary superpositions of the left and right moving sectors. On the other hand, in sLQC the expectation values of volume diverge both in asymptotic future and past $(\phi \rightarrow \pm \infty)$.
For any arbitrary state in the physical Hilbert space, $\langle \hat V|_\phi \rangle$ has a minimum value $V_{\mathrm{min}} = 2 \sqrt{V_+ V_-}/{||\chi||^2}$
which is reached at the bounce time $\phi_B = (2 \sqrt{12 \pi G})^{-1} \ln(V_+/V_-)$. A consistent histories analysis in sLQC yields the quantum probability for
the bounce to be unity \cite{consistent3}. Unlike the Wheeler-DeWitt theory where big bang and big crunch singularities are inevitable, in sLQC these singularities are resolved for the generic states.

The fluctuations of the volume and the momentum observable can be computed in a similar way, which give important insights on the evolution of the states across the bounce.
This issue is tied to understanding the way detailed properties of the universe in sLQC post-bounce branch are influenced by the initial 
state in the pre-bounce branch (or vice versa). Using sLQC constraints on the growth of the relative fluctuations have been obtained which 
show that a state which is semi-classical at very early times before the bounce retains its semi-classical properties after the bounce 
\cite{cs2,kp2,cm1,cm2,tp-memory}. In particular, triangle inequalities relating the relative fluctuations of volume and momentum provide strong 
constraints on degree by which the fluctuations can change across the bounce \cite{kp2}. 
These inequalities have been recently rigorously tested using extensive numerical simulations for semi-classical states as well as for 
states which are not  sharply peaked \cite{dgs2,squeezed}. These inequalities are found to remain valid for all the numerical simulations 
performed till date.

Useful insights on the details of the singularity resolution in sLQC emerge on analyzing expectation values of the energy density of massless scalar field, whose
corresponding observable is
\be \hat \rho|_\phi = \f{1}{2}\, (\hat A|_\phi)^2 ~~~\,
{\mathrm{where}} \, ~~~\hat A|_\phi = (\hat V|_\phi)^{-1/2} \, \hat
p_\phi  (\hat V|_\phi)^{-1/2} ~. \ee
The expectation values $\langle \hat \rho|_\phi \rangle$ computed at some $\phi = \phi_o$ is,
\be \langle \hat \rho|_{\phi_o} \rangle
\, = \, \f{3}{8 \pi G \gamma^2} \, \f{1}{\lambda^2}
\f{\left(\int_{-\infty}^\infty \d x |\p_x F|^2\right)^2}{\left(\int_{-\infty}^\infty \dd
x |\p_x F|^2 \cosh(\sqrt{12 \pi G} x)\right)^2}\,  \ee
which is bounded above by $\rho_{\mathrm{sup}}$:
\be \label{rhosup}
\rho_{\mathrm{sup}} = \f{3}{8 \pi \gamma^2 G \lambda^2} =
\f{\sqrt{3}}{32 \pi^2 \gamma^3 G^2 \hbar} \approx 0.41
\rho_{\mathrm{Pl}} ~. \ee
This value is in excellent agreement with the value of energy density at the bounce
obtained using numerical simulations for the states which are sharply peaked at the
classical trajectory at late times \cite{aps3,dgs2}. As pointed out earlier, for the states which
are widely spread, the value of energy density at the bounce in general turns out to be
less than the above value \cite{dgs2}. The same conclusion holds true for the
states which are squeezed \cite{cm2,squeezed} or for states with more complex waveforms \cite{squeezed}.
Extensive numerical simulations for various kinds of states have shown that the above supremum of the energy
density always holds true \cite{dgs2,squeezed}.

The generic bound on the energy density is a direct consequence of the quantum geometry which manifests through the area gap $\lambda^2$. It is also related to the ultra-violet cutoff for the eigenfunctions of the evolution operator which decay exponentially below the volume at which this energy density is reached. The evidence of this feature was first found numerically \cite{aps2,aps3}, which has been recently rigorously confirmed using sLQC \cite{craig}.  If the area gap is put to zero, the maximum of the energy density
becomes infinity and the ultraviolet cutoff on the eigenfunctions disappears. 
 Note that sLQC can be approximated to Wheeler-DeWitt theory for any given accuracy $\epsilon$ in a semi-infinite interval of time $\phi$ 
by appropriately shrinking the area gap. However, this is not possible if the entire infinite range of $\phi$ is considered. Then 
irrespective of the choice of a finite area gap, the differences between sLQC and Wheeler-DeWitt become arbitrarily large  in some range of 
time $\phi$. In this sense of global time evolution, the Wheeler-DeWitt theory is not a limiting case of  
sLQC. It turns out that sLQC is a fundamentally discrete theory and the limit $\lambda \rightarrow 0$ does not 
lead to a continuum theory. This feature of sLQC is not shared
by the examples in polymer quantum mechanics where the continuum limit exists in the limit when the discreteness parameter vanishes \cite{afw,cvz2}.

To summarize, sLQC has played an important role in proving the robustness of results on the bounce that were first observed in numerical simulations within LQC. The exact solvability of this model provides many insights on the supremum of the expectation value of the energy density observable, bounds on the growth of fluctuations across the bounce, and relation with the Wheeler-DeWitt theory.

\section{LQC in more general cosmological spacetimes}\label{sec3}
In the previous section, we discussed the way quantum geometric effects in LQC resolve the classical big bang/big crunch singularity in the $k=0$
isotropic and homogeneous spacetime, and result in a quantum bounce of the universe near the Planck scale. This result opens a new avenue to explore and develop novel non-singular paradigms in the very early universe, and to answer fundamental questions related to the structure of the spacetime near the singularities. For this it is necessary to
extend the results on singularity resolution in LQC to more general settings. The goal of this section is to summarize the main developments in these directions. In Sec. 3.1, we start with a brief discussion on the generalization of the bounce results in different isotropic models -- with spatial curvature, cosmological constant and an  inflationary potential, focusing in particular on the properties of the quantum evolution operator and subtle features of the loop quantization.\footnote{Loop quantization of isotropic model has also been performed in the presence of radiation. For details, we refer the reader to Ref. \cite{rad}.}  This is followed by a discussion of two interesting applications in effective dynamics. 
Sec. 3.2 deals with the loop quantization of Bianchi models, where aspects of quantum theory and effective dynamics of Bianchi-I model is discussed in some detail. In Sec. 3.3, we discuss the application of LQC techniques to the Gowdy models
which have provided useful insights on the singularity resolution in the presence of inhomogeneities.

\subsection{Quantization of other isotropic models}
In the following, in Sec. 3.1.1 till Sec. 3.1.3, we summarize some of the main features of the isotropic models in LQC which have been quantized using the procedure outlined in Sec. 2.2. In all of these homogeneous models, one starts with the gravitational part of the classical Hamiltonian constraint in terms of the fluxes and the field strength of the connection, and expresses them in terms of the triads and the holonomies computed over a closed loop whose minimum area is given by $\lambda^2 = \Delta \lp^2$. The edge lengths of the holonomies $\bar \mu$ are functions of triads given by eq. (\ref{mubar}).  The elements of holonomies form an algebra of almost periodic functions, and their action on the states in the volume representation is by uniform translations. As in the case of the $k=0$ model, one obtains a quantum difference equation with uniform discreteness in volume. The scalar field plays the role of time in the quantum evolution, and one can introduce an inner product and a family of Dirac
observables to extract physical predictions.
Extensive numerical simulations confirm the existence of bounce which occurs at $\rho \approx 0.41 \rcr$ for the sharply peaked states. Effective dynamics turns out to be
in excellent agreement with the underlying loop quantum dynamics in all of these models. The last part of this section exhibits two applications of effective dynamics, where we
discuss the way effective spacetime description provides important insights on the generic resolution of singularities and the naturalness of inflation in LQC.

\subsubsection{Spatially closed model:}
The isotropic and homogeneous $k=1$ model with a massless scalar field provides a very useful stage to carry out precise tests on the ultra-violet and infra-red limits in LQC. This is because in the classical theory, the scale factor in $k=1$ model recollapses at a value determined by the momentum $p_{(\phi)}$. In the past of the classical evolution, the universe encounters a big bang singularity, and in the future it encounters a big crunch singularity. A non-singular quantum cosmological model should not only resolve both of
the past and the future singularities, but must also lead to recollapse at the scales determined by the classical theory. Using the earlier quantization in LQC, Green and Unruh found that though the singularity is resolved, one is not able to obtain recollapse at the large scales predicted by the classial theory \cite{gu}. This limitation was tied to the unavailability of the inner product and detailed knowledge of the properties of the quantum evolution operator in the earlier works. These limitations were overcome in the
loop quantization of the $k=1$ model with a massless scalar field following the quntization procedure in the $k=0$ model outlined in Sec. 2.2 \cite{apsv,warsaw1}.\footnote{The model has been recently quantized in a different way following the same strategy \cite{ck2}.} The resulting
quantum Hamiltonian constraint, for lapse chosen to be $N=a^3$, takes the following form:
\ba \label{k1-hc}
\p_\phi^2 \Psi(\nu,\phi) \, &=& \, \nonumber - \Theta_{(k=1)} \, \Psi(\nu,\phi)\\
&=& - \nonumber \Theta \Psi(\nu,\phi) + \f{3 \pi G}{\lambda^2} \,
\nu \bigg[\sin^2\left(\f{\lambda}{\tilde K \nu^{1/3}} \f{\ell_o}{2}
\right) \, \nu \, \\
&& ~~~~~~~~~~~~~~~~~~~~~~~ - \, (1 + \gamma^2) \left(\f{\lambda}{\tilde K}
\f{\ell_o}{2}\right)^2 \, \nu^{1/3}\bigg] \, \Psi(\nu,\phi)  
\ea
where $\Theta_{(k=1)}$ is a positive definite, self adjoint operator. In contrast to the quantum evolution operator $\Theta$ of the $k=0$ model, $\Theta_{(k=1)}$ has a discrete
spectrum. This property is tied to the behavior of the eigenfunctions of $\Theta_{(k=1)}$ which decay exponentially for volumes greater than the recollapse volume in the classical theory, and also below a particular value of volume when the spacetime curvature reaches Planck scale. Numerical simulations with sharply peaked Gaussian states, and analysis of Dirac observables $\hat p_{(\phi)}$ and $\hat V|_\phi$,  show that the $k=1$ loop quantum universe bounces at the volume where the exponential decay of the eigenfunctions occurs in the Planck regime, with a maximum in the expectation values of energy density observable given by $0.41 \rcr$. The loop quantized model also recollapses at the value in  excellent agreement with the classical theory. States preserve their peakedness properties through bounces and recollapses, and
the quantum evolution continues forever by avoiding the big bang and big crunch singularities, providing a non-singular cyclic model of the universe. The effective dynamics
obtained from an effective Hamiltonian constraint with a form similar to (\ref{effham1}) provides an excellent agreement to the loop quantum dynamics at all the scales. The loop quantization of $k=1$ model successfully demonstrates that in LQC not only are the classical singularities resolved, but the theory also agrees with GR with an extra-ordinary precision at classical scales.

\subsubsection{Positive and negative cosmological constant:}
The case of  positive cosmological constant $\Lambda$, with a massless scalar field in the spatially flat isotropic and homogeneous spacetime is interesting due to various conceptual and phenomenolgical reasons. In the classial theory, the model has a big bang singularity in the past, and undergoes accelerated expansion in the future evolution. The universe expands to an infinite volume in an infinite proper time $t$, but at a  finite value of the scalar field $\phi$. Thus, in the relational dynamics with $\phi$ as a clock, the Hamiltonian vector field on the phase space is incomplete. With an analytical extension, the dynamical trajectories in the classical theory start from a big bang singularity at $\phi = -\infty$, and encounter a big crunch at $\phi = \infty$ \cite{ap}. In terms of the time $\phi$, the classical evolution thus turns out to be much richer. Following the startegy for the loop quantization of $k=0$ FLRW model with a massless scalar, loop quantization can be performed rigorously, which
leads to the following evolution equation \cite{ap}:
\be \label{qhc-lambda} \p_\phi^2\,\Psi(\nu,\phi) =
-\Theta_{\Lambda_+}\, \Psi(\nu,\phi) := -\Theta\, \Psi(\nu,\phi) +
\f{\pi G\gamma^2\,\Lambda}{2}\, \nu^2\,\Psi(\nu,\phi)\, ,  \ee
where $\Lambda > 0$. The operator $\Theta_{\Lambda_+}$ is not essentially self-adjoint, and one needs to find its self-adjoint extensions \cite{kp1,ap}. For any choice of such an extension,
the spectrum of $\Theta_{\Lambda_+}$ is discrete. It turns out that the details of the physics are independent of the choice of the extension for large eigenvalues of $\Theta_{\Lambda_+}$. Numerical simulations with sharply peaked states show the existence of bounce when the energy density of massless scalar field and cosmological constant
becomes approximately equal to $0.41 \rho_{\rm{Pl}}$. Interestingly, for $\phi$ as a clock, infinite volume is reached in finite $\phi$. In the quantum theory, evolution continues beyond this point in relational time $\phi$ and results in a contracting trajectory.   In this way, the evolution in this model mimics the cyclic universe with $\phi$ as time.

The loop quantization of $k=0$ model with a massless scalar field and a negative cosmological constant was first discussed briefly in Ref.\cite{aps3}, and studied in detail in Ref. \cite{bp}. As in the $k=1$ model, the classical universe has a big bang singularity in the past, and a big crunch singularity in the future after the negative cosmological constant results in a recollapse in the expanding branch. In LQC, the quantum Hamiltonian constraint turns out to be,
\be \label{qhc-prime-lambda} \p_\phi^2\,\Psi(\nu,\phi) =
-\Theta_{{\Lambda_{-}}}\, \Psi(\nu,\phi) := -\Theta\,
\Psi(\nu,\phi) - \f{\pi G\gamma^2\, |\Lambda|}{2}\,
\nu^2\,\Psi(\nu,\phi) ~. \ee
The operator $\Theta_{{\Lambda_{-}}}$ is essentially self-adjoint with a discrete spectrum. At large eigenvalues, the spacing between the eignvalues is nearly uniform. Sharply peaked states constructed with such eigenvalues undergo nearly cyclic evolution in LQC, avoiding the big bang singularity in the past and the big crunch singularity in the future, with quantum bounces occuring at $\rho \approx 0.41 \rho_{\rm{Pl}}$. As in the $k=1$ model, the universe recollapses at the volume predicted by the classical theory.

\subsubsection{Inflationary potential:}
In the classical theory, inflationary spacetimes are past incomplete \cite{bgv}. A natural question is whether in LQC,  one can construct non-singular inflationary models. The problem is challenging because of several reasons. Unlike the models considered so far, in the presence of a potential, $p_{(\phi)}$ is not a constant of motion and therefore $\phi$ does not serve as a global clock. However, $\phi$ can still be used as a local clock in  portions of the pre-inflationary epoch where it has a monotonic behavior. For the $\frac{1}{2}m^2 \phi^2$ potential, the loop quantization leads to the  following quantum evolution equation \cite{aps4},
\be \label{qhc-inflaton} \p_\phi^2\,\Psi(\nu,\phi) = -\Theta_{(m)}\,
\Psi(\nu,\phi) := -\big(\Theta -4 \pi G\gamma^2\,  m^2\phi^2\, \nu^2\big)\,\,\Psi(\nu,\phi)\, . \ee
Note that since $\Theta_{(m)}$ depends on time $\phi$, obtaining the inner product becomes more subtle. The operator $\Theta_{(m)}$ is equivalent to $\Theta_{(\Lambda_+)}$ in (\ref{qhc-lambda}) for any fixed value of $\phi$, and hence fails to be essentially self-adjoint. For each value of $\phi$, $\Theta_{(m)}$ admits self-adoint extensions.
The physical Hilbert space can be obtained given a choice of these extensions. Numerical simulations with sharply peaked states show that the quantum evolution resolves the past singularity and results in a quantum bounce when the energy density of the inflaton field reaches $0.41 \rho_{\rm{Pl}}$ \cite{aps4}. Further, the classical GR trajectory is recovered when the spacetime curvature becomes much smaller than the Planck value. Thus, loop quantum gravitational effects make inflation past complete.  A detailed understanding of the physics of this model and its relation to the choice of self-adjoint extensions is an open issue.

\subsubsection{Some applications of the effective dynamics:}
We now discuss two applications of the effective dynamics in the isotropic model. The first example probes the question whether singularities are generically
resolved in LQC, and under what conditions, and the second example deals  with the naturalness of the inflationary scenario in LQC.\\

\noindent
$\bullet$ {\it{Generic resolution of  strong singularities:}} %Existence of bounce in the loop quantization of the isotropic models provides an evidence that quantum geometric effects  can successfully resolve the classical singularities.
Apart from the resolution of the big bang/big crunch singularities in the quantum theory, the resolution of various other types of singularities such the big rip singularity has been achieved in the effective spacetime description in LQC \cite{sst,ps09,ps-fv,gumju,naskar}. An important question is whether quantum geometric effects resolve all the spacelike singularities, or are there certain types of singularities which are not resolved. Here it is to be noted that even in GR, not all singularities are harmful. Certain singularities even if characterized by divergences in the components of spacetime curvature, are harmless because the tidal forces turn out to be finite \cite{ellis,tipler,krolak}. It turns out that such  singularities -- known as weak singularities, are not resolved
in LQC \cite{portsmouth,ps09}. Such singularities are tied to the divergence in the spacetime curvature caused by the divergence in the pressure even when the energy density is bounded above. Note that for various LQC spacetimes, including isotropic, anisotropic and certain Gowdy models, expansion scalar and anisotropic shear are bounded \cite{cs3,pswe,gs1,ks-bound,ck-bianchi9,gowdy}. However,  it is straightforward to see that LQC allows divergence of curvature invariants. A straightforward computation of  Ricci scalar $R$ from the
modified Friedman (\ref{fried}) and the Raychaudhuri equation (\ref{raych}), shows that 
% \be \label{Ricci} R = 6 \left(H^2 + \f{\ddot a}{a} \right) =  8
% \pi G \rho \, \left(1 + \f{2 \rho}{\rcr}\right) - 24 \pi G P
% \left(1 - \f{2 \rho}{\rcr} \right) ~.
% \ee
if the equation of state of matter is such that $P(\rho) \rightarrow \pm \infty$ at a finite value of $\rho$, then the Ricci scalar diverges. However, geodesics can be extended beyond all such events \cite{ps09}.
On the other hand, strong singularities -- the ones for which tidal forces are infinite, are generically resolved for matter with arbitrary equation of state in the $k=0$ isotropic and homogeneous model in LQC \cite{ps09}. The expansion scalar of the geodesics remains bounded in the effective spacetime which turns out to be geodesically complete.
 For the spatially curved model, a phenomenological analysis of effective dynamics confirms that strong singularities are resolved where as weak singularities are ignored by the quanutm geometric effects \cite{ps-fv}. The results on generic resolution of strong singularities have also been generalized in the presence of anisotropies in the Bianchi-I model \cite{ps11,ps15} and Kantowski-Sachs spacetime \cite{ss1}. These results provide a strong indication that resolution of singularities may be a very generic phenomena due to the quantum gravitational effects in LQC. A fundamental question is whether these results point towards a non-singularity theorem. Future work in this direction is expected to  reveal an answer to this important question.\\

\noindent
$\bullet$ {\it{Probability for inflation:}}  The inflationary paradigm has been extremely successful in providing a description of the early universe in the FLRW model. It is
natural to ask whether it can be successfully embedded in LQC. Various inflationary models have been considered in LQC using the modified Friedman dynamics, including single field inflation with $m^2 \phi^2$ potential \cite{svv,as,as3,ck-inflation,barrau_infl1}, multi-field inflationarty models \cite{assisted}, tachyonic inflation \cite{tachyonic},
with non-minimally coupled scalar fields \cite{nonminimal}, and even in the presence of anisotropies  \cite{gs3}. An important question in this setting is, if
%question is whether the inflationary scenario can be embedded or realized within LQC in a natural way. That is, consider the matter content is given by a scalar field with a suitable potential, e.g.\ the simple quadratic potential $V(\phi)=1/2\, m^2\phi^2$ often used in inflation. If now
we solve the spacetime dynamics in LQC starting the evolution at the Planck era of the universe, does a phase of slow-roll inflation compatible with observations appear at some time in the future evolution? If so, does this happen  for generic initial conditions or only for very specific, fine-tuned values of the initial data? In the context of GR, it has long been argued that inflationary trajectories are {\em attractors} in the space of solutions. Similar conclusions have been found in LQC \cite{svv}, but existence
of attractors does not tell us about the probabilities unless a suitable measure is defined.  A detailed analysis of this questions was performed in Refs. \cite{as,as3} in a spatially flat FLRW background. The authors of this reference proceed by computing the {\em fractional volume} in the space of solutions  occupied by physical trajectories which the desired properties  ---an inflationary phase compatible with observation at some time in the evolution. But as they point out, the presence of gauge degrees of freedom makes the construction of the measure needed to compute those volumes quite subtle. The natural strategy, namely the projection of the natural Liouville measure to the reduced, or gauge fixed space of solutions,  produces ambiguous results ---it turns out, as pointed out in \cite{ck-inflation}, that this fact is the origin of disparate results in the literature \cite{klm, gt}. The ambiguity, on the other hand, can be resolved by introducing a preferred moment in time in the evolution. In GR, 
there is no
such a preferred instant, but in LQC the existence of the bounce that {\em every} trajectory experiences provides the required structure. The authors of Refs. \cite{as,as3} use this fact to show that, by  assuming a flat probability distribution in the space of initial conditions for matter and geometry at the bounce time, a $99.9997\%$ of the volume of that space corresponds to solutions that will encounter observationally favored inflation during the evolution. The conclusion is therefore that the inflationary  attractor is also present in LQC. See Ref. \cite{corichi-sloan} for further explanation of the underlying reason of this attractor mechanism.

\subsection{Bianchi-I model}\label{secBianchi}
The Bianchi-I model is one of the simplest settings to understand the way anisotropies in the spacetime play an
important role on the physics near the classical singularities. Due to an interplay of the Ricci and Weyl components of the spacetime curvature, the structure of
the singularities is very rich in comparison to the isotropic models. The spacetime metric of the Bianchi-I model is given by
\be \dd s^2 = - \dd t^2 + a_1^2 \dd x_1^2 + a_2^2 \dd x_2^2 +
a_3^2 \dd x_3^2, \ee
where $a_i$ denote the directional scale factors. Unlike the isotropic models,
where the big bang singularity is characterized by the universe shrinking to a point of zero scale factor, the
big bang singularity in the anisotropic models in GR can be of the shape of a cigar, a pancake, a barrel or a point depending on the behavior of the directional scale factors, captured by the Kasner exponents $k_i$, defined via $a_i \propto t^{k_i}$. It turns out that in general, unless one considers matter which has an equation of state, given by the ratio of the pressure and the energy density, to be greater than or equal to unity, the approach to the singularity is dictated by the anisotropic shear.  In more general situations, such as in the Bianchi-IX models where the presence of spatial curvature leads to even richer dynamics, the approach to the singularity is oscillatory in the Kasner exponents and leads to the mixmaster behavior \cite{berger}. According to the BKL conjecture, in the inhomogeneous spacetimes the approach to the  singularity is such that the spatial derivatives can be ignored in comparison to the time
derivatives, and each point of the space asymptotically behaves as in the Bianchi-IX model \cite{bkl}. Thus, understanding the way singularity resolution occurs in Bianchi models is very important to understand singularity resolution in general.

In LQC, a quantization of Bianchi-I \cite{awe2}, Bianchi-II \cite{awe3} and Bianchi-IX models \cite{we,pswe} has been performed which leads to a non-singular quantum constraint equation.
However, physical implications in the quantum theory in terms of the expectation values of Dirac observables have only been studied for the Bianchi-I vacuum model \cite{madrid-bianchi} for an earlier quantization \cite{chiou_bianchi,cv}.\footnote{This quantization has some limitations related to the dependence of physical predictions on the shape of the fiducial cell which is  introduced to define symplectic structure \cite{awe2,cs3}.  Nevertheless, it is consistent when the spatial topology is a 3-torus and provides important insights on the physics at the Planck scale and the nature of bounce in Bianchi-I model.} Various interesting results on the physics of the Bianchi models have been obtained using effective dynamics, which include existence of Kasner transitions
across the bounce in Bianchi-I model\cite{cs2}, existence of inflationary attractors \cite{cs3}, constructing non-singular cyclic models in the Ekpyrotic scenario \cite{csv}, generic bounds on geometric scalars \cite{cs3} and the resolution of strong curvature singularities \cite{ps11,ps15}. In the following, we outline the
quantization of the Bianchi-I model with a massless scalar fiels as performed in Ref. \cite{awe2}, and discuss some of the features of the effective dynamics. For details of the loop quantization and the resulting physics of Bianchi-II and Bianchi-IX spacetimes, we refer the readers to the original works \cite{cv,awe3,we,pswe,ac-bianchi,gs1,CKM,cm-b9,
ck-bianchi9}. 

Utilizing the symmetries of the spatial manifold, which as in the $k=0$ isotropic model can be of $\mathbb{R}^3$ or $\mathbb{T}^3$ topology, the Ashtekar-Barbero connection and
the densitized triad in the Bianchi-I model can be written as
\be\label{sym_red} A^i_a \, = \, c^i (L^i)^{-1} \w^i_a, ~~
\mathrm{and} ~~ E^a_i = p_i L_i V_o^{-1} \sqrt{\q}~ \e^a_i ~,
\ee
where $c^i$ and $p_i$ are the symmetry reduced connections and triads and $L_i$ denote the coordinate lengths of the fiducial cell in the case of $\mathbb{R}^3$ topology. For the $\mathbb{T}^3$ topology, one does not need to introduce a fiducial cell and $L_i$ can be set to $2 \pi$. Note that the coordinate volume $V_o = L_1 L_2 L_3$ with respect to the fiducial metric $\q_{ab}$ changes if the individual $L_i$ are rescaled. In the classical theory, physics is invariant under the change in the rescalings in $L_i$. This also turns out to be true for the loop quantization of the Bianchi-I model discussed here.\footnote{In the earlier quantization prescription \cite{chiou_bianchi,madrid-bianchi},  the resulting physics is not invariant under the change in shape of the fiducial cell if the topology is non-compact \cite{awe2,cs3}.} The triad components are related to the directional scale factors as
\be\label{triad_sf} p_1 \, = \, \varepsilon_1\, L_2 L_3 \, |a_2
a_3|, ~~ p_2 \,= \, \varepsilon_2 L_1 L_3 \, |a_1 a_3|, ~~ p_3 \, =
\, \varepsilon_3  L_1 L_2 \, |a_1 a_2| ~ , \ee
where $\varepsilon_i = \pm 1$ depending on the triad orientation. For the massless scalar field, the classical Hamiltonian constraint in terms of the symmetry reduced variables for lapse $N = a_1 a_2 a_3$, is given by
\be \label{clH} {C}_H = -\, \f{1}{8 \pi
G \gamma^2}{(c_1 p_1 \, c_2 p_2 + c_3 p_3 \, c_1 p_1 + c_2 p_2 \,
c_3 p_3)} \, + \f{p_{(\phi)}^2}{2}\, \, \approx \,\, 0\, . \ee
Using Hamilton's equations, one finds that
\be\label{cipi}
c_i p_i - c_j p_j = V(H_i - H_j) = \gamma \kappa_{\ij} ~,
\ee
where $\kappa_{\ij}$ is a constant antisymmetric matrix, and $H_i$ denote the directional Hubble rates $H_i = \dot a_i/a_i$.

To understand the dynamical evolution, it is useful to introduce the  mean Hubble rate and the shear scalar in this model. These are
{\it{kinematically}} obtained from the trace and the symmetric tracefree parts of the expansion tensor defined as the covariant derivative of the timelike vector field tangential to the geodesics. The mean Hubble rate and the shear scalar
in the Bianchi-I spacetime are:
\be
H = \f{1}{3}(H_1 + H_2 + H_3)~, ~~~ \mathrm{and} ~~~ \sigma^2 = \f{1}{3} \, \left((H_1 - H_2)^2 + (H_2 - H_3)^2 + (H_3 - H_1)^2\right) ~,
\ee
where $H = \dot a/a$ with $a = (a_1 a_2 a_3)^{1/3}$.

In the classical theory, using eq.(\ref{cipi}), the shear $\Sigma^2 := \sigma^2 V^2/6$ turns out to be a constant.\footnote{If matter has a non-vanishing anisotropic stress, as in the case of magnetic fields, $\Sigma^2$ is not constant in the classical theory. For a phenomenological investigation of the Bianchi-I model in LQC in such a situation, see Ref. \cite{roy}.}  The Hamilton's equations for the triads, yield the generalized Friedman equation:
\be\label{gen_fried_class} H^2 = \f{8 \pi G}{3} \rho \, + \,
\f{\Sigma^2}{a^6} ~. \ee
At the classical big bang singularity, the mean Hubble rate, energy density of the scalar field $\rho$ and the shear scalar $\sigma^2$ diverge, and the geodesic evolution breaks down. Note that the shear scalar $\sigma^2 \propto a^{-6}$, and thus it diverges at the same rate as the massless scalar field energy density. In the presence of other matter sources such as dust and radiation, or an inflationary potential, since the energy density diverges slower than $a^{-6}$, the shear scalar dominates  near the classical singularity and the singularity is necessarily  anisotropic.

Let us now summarize the loop quantization of the Bianchi-I model. It is carried out with a similar procedure as in isotropic model, where one starts with a Hamiltonian constraint expressed in
terms of triads $E^a_i$ and the field strength of the holonomies $F_{ab}^{~~i}$. The holonomies yield an algebra of almost periodic functions of connections $c_i$, and the
field strength can be computed by considering holonomies over a square loop $\Box_{ij}$. Due to the presence of anisotropies, the relation between the edge lengths $\bar \mu_i$ of the loops turns out to be \cite{awe2}:
\be \label{bianchi_mub1} \bar \mu_1 = \lam \sqrt{\f{ |p_1|
}{|p_2 p_3|}}, ~~~~ \bar \mu_2 = \lam \sqrt{\f{|p_2|}{|p_1
p_3|}}, ~~~ \mathrm{and} ~~~\bar \mu_3 = \lam
\sqrt{\f{|p_3|}{|p_1 p_2|}} ~, \ee
which reduces to the isotropic relation (\ref{mubar}) when $p_1 = p_2 = p_3$. Due to the functional dependence on direction triads, the action of the elements of the holonomy algebra $\exp(i \bar \mu_i c^i)$
is very complicated on the states $\Psi(p_1,p_2,p_3)$. It is more convenient to work with states $\Psi(l_1,l_2,v)$ where $l_i$ are defined via $p_i =
({\rm sgn}\, l_i)\,\,(4 \pi \gamma \lambda \lp^2)^{2/3}\,\, l_i^2$, and $v = 2 (l_1 l_2 l_3)$. Computing the action of the field strength and the triad operators on these states, which also required to be symmetric under the change of the orientations of the triad, the quantum Hamiltonian constraint turns out to be
\be \label{bianchi-qconst}
\p_\phi^2 \Psi(l_1,l_2,v;\phi) = - \Theta_{\mathrm{B-I}}\, \Psi(l_1,l_2,v;\phi)
\ee
where
 \ba \Theta_{\mathrm{B-I}}\, \Psi(l_1,l_2,v;\phi) &=& \f{\pi G
 \hbar^2}{8} \, \sqrt{v} \, \bigg[(v + 2) \sqrt{(v + 4)}
 \Psi_4^+(l_1,l_2,v) \, - \, (v+2) \sqrt{v} \Psi_0^+(l_1,l_2,v;\phi)\nonumber \\
 && - (v-2) \sqrt{v} \Psi_0^-(l_1,l_2,v;\phi) + (v-2) \sqrt{|v - 4|}
 \Psi_4^-(l_1,l_2,v;\phi)\bigg] ~.\ea
Here $\Psi^{\pm}_4$ and $\Psi^{\pm}_0$ are defined as
 \ba \label{Psi4} \Psi_4^\pm\lvp &=& \nonumber \Psi\left(\vfu \lu,
 \vfd \ld, v
 \pm 4\right) \, + \, \Psi\left(\vfu \lu, \ld, v \pm 4\right) \\
 && \nonumber + ~ \Psi\left(\vfd \lu, \vfu \ld, v \pm 4\right) \,
 + \, \Psi\left(\vfd \lu, \ld, v \pm 4\right) \\
 && + ~ \Psi\left(\lu, \vfd \ld, v \pm 4\right) \, + \,
 \Psi\left(\lu, \vfu \ld, v \pm 4\right)  ~, \ea
% %
 and
% %
 \ba \label{Psi0} \Psi_0^\pm\lvp &=& \nonumber \Psi\left(\vfd \lu,
 \vft \ld, v\right) \,
 + \, \Psi\left(\vfd \lu, \ld, v\right) \\
 && \nonumber + ~ \Psi\left(\vft \lu, \vfd \ld, v\right) \, +
 \, \Psi\left(\vft \lu, \ld, v\right) \\
 && + ~ \Psi\left(\lu, \vft \ld, v\right) \, + \, \Psi\left(\lu, \vfd
 \ld, v\right)  ~. \ea
 An important property of the above quantum difference equation is the following. If one starts with an initial wavefunction which is peaked on a non-zero volume and
 which vanishes at $v=0$, then in the quantum evolution it can not have support on the zero volume. The classical singularity at $v=0$ is decoupled from the evolution in the quantum theory \cite{awe2}. This shows that for such wavefunctions, the spacetime curvature will remain bounded throughout the physical evolution. Further, it can be shown that the isotropic LQC for $k=0$ model is recovered by integrating out anisotropies.  To explore the physics in detail,
 numerical simulations on the lines of the isotrpic models are required to compute expectation values of the Dirac observables, which in this case are $\hat V|_\phi, \hat l_1|_\phi$ and $\hat l_2|_\phi$. Since the form of the quantum evolution operator (\ref{bianchi-qconst}) is quite complicated in comparison to the isotropic quantum evolution operator (\ref{thetacons}), numerical simulations are technically difficult. However, insights have been gained on simplifying the quantum constraint to obtain physical solutions \cite{hender-pawlowski}.

Useful insights on the Planck scale physics in the Bianchi-I model in LQC has been gained using the effective Hamiltonian constraint, which following the derivation in the isotropic
model in LQC, is given by \cite{cv,awe2}:
 \be\label{effham_bianchi} {C}^{\,\mathrm{eff}}_H =  - ~\f{1}{8
\pi G \gamma^2 (p_1 p_2 p_3)^{1/2}} \left(\f{\sin(\bar \mu_1
c_1)}{\bar \mu_1} \f{\sin(\bar \mu_2 c_2)}{\bar \mu_2} p_1 p_2 +
\mathrm{cyclic}
 ~~ \mathrm{terms}\right) ~~ + ~~ {\cal H}_{\mathrm{matt}}~  \\
\ee
where ${\cal H}_{\mathrm{matt}}$ denotes the matter Hamiltonian. The expression for the energy density can be obtained from the vanishing of the effective Hamiltonian constraint,
${C}^{\,\mathrm{eff}}_H \approx 0$, and it turns out to be
\be\label{density_bianchi1_eff} \rho = \f{1}{8 \pi G \gamma^2
\lam^2} \left(\sin(\bar \mu_1 c_1) \, \sin(\bar \mu_2 c_2) +
\mathrm{cyclic} ~~ \mathrm{terms}\right) ~. \ee
The energy density is bounded and has an absolute maximum $\rcr \approx 0.41 \rho_{{\rm{Pl}}}$, same as in the isotropic $k=0$ model.   Using Hamilton's equations,
one can compute the mean Hubble rate and the shear scalar which also turn out to be bounded \cite{cs3,gs1}: $H_{\rm{max}} = 1/(2 \gamma \lambda)$ and $\sigma^2_{\rm{max}} =  10.125/(3 \gamma^2 \lambda^2)$. Similar bounds have been found in the presence of spatial curvature, in Bianchi-II and Bianchi-IX models \cite{gs1,pswe,ck-bianchi9}. It has been so far difficult to find a consistent modified generalized Friedmann equation in terms of the energy density $\rho$ and anisotropic shear scalar $\sigma^2$, except when anisotropies are weak  \cite{cv}.\footnote{One can rearrange the Hamilton's equations to obtain an equation for the mean Hubble rate which mimics the classical generalized Friedmann equation exactly by defining a `quantum shear' \cite{barrau_aniso}. However, the limitation with such an approach is that the `quantum shear' does not consistently capture the anisotropic shear in the effective spacetime, and one loses important information about the way anisotropies influence the
Planck scale phenomena in LQC.} However, using Hamilton's equations the effective dynamics has been explored in a lot of detail.  The effective dynamics of the Bianchi-I model turns out to be non-singular for various types of matter resulting in a bounce of the mean scale factor at the Planck scale. For perfect fluids  with a vanishing anisotropic stress and  with an arbitrary equation of state greater than $-1$, the curvature invariants are bounded and all strong curvature singularities are generically resolved \cite{ps11,ps15}.
The approach to the bounce can be characterized using Kasner exponents and one finds that depending on the ratio of the intial anisotropy and energy density,  the bounce can be associated to a barrel, cigar, pancake or a point like structure. Interestingly, the structures before and after the bounce can change in general and follow certain selection rules \cite{gs2}. Depending on the anisotropic parameters,  some transitions are also completely forbidden. Note that such Kasner transitions are absent in the Bianchi-I model in GR, and are so far known to arise only in LQC. 

To summarize, the Bianchi-I model with a massless scalar field can be quantized in LQC in a similar way as the isotropic models, and the resulting quantum Hamiltonian constraint turns out to be a non-singular quantum difference equation. Unlike in the classical theory, the energy density, mean Hubble rate and anisotropies remain bounded throughout the evolution. The physics of the quantum Bianchi-I spacetime is
considerably richer than the isotropic model and has provided a robust picture of the Planck scale physics in LQC. There are several interesting avenues to explore, including the way quantum geometry affects Mixmaster behavior which is expected to give important insights on the resolution of singularities in more general situations. Thanks to
the formulation of the BKL conjecture in the connection variables \cite{ahs}, a stage is set to carry out a rigorous comparison between the quantum and the classical description of spacetimes in the Bianchi models, and to gain valuable insights on the generic resolution of singularities.

\subsection{Gowdy models and the Hybrid quantization}
So far we have discussed spacetimes with a finite numer of degrees of freedom. A long standing issue in quantum cosmology is whether the
physical implications obtained in the mini-superpsace setting can be trusted for spacetimes with an infinite number of degrees of freedom.
One of the directions which has been explored  to go beyond the assumption of homogeneity is the Gowdy midi-superspace
spacetimes which  have been quantized using a hybrid method in which the homogeneous modes are loop
quantized, and the inhomogeneous modes are Fock quantized \cite{hybrid1,hybrid2,hybrid3,hybrid4}. The results of singularity resolution
obtained in homogeneous models are found to be robust in the Gowdy models for vacuum \cite {hybrid3} and
also in presence of a massless scalar field \cite{hybrid4}. In the following,
we outline the main features of this approach for the vacuum case.

The Gowdy spacetimes which have been studied so far in LQC are the one with linear polarization. The spatial manifold is  $\mathbb{T}^3$  coordinatized by  $(\theta,\sigma,\delta)$. The spacetime has two Killing fields, $\p_\sigma$ and $\p_\delta$, which are hypersurface
orthogonal. The spatial dependence in the fields is captured completely by $\theta$. Using the underlying symetries, a partial gauge
fixing can be performed and  one is left with two global
constraints: the diffeomorphism constraint  $C_\theta$ and the Hamiltonian constraint $C_H$. The metric components are periodic in
$\theta$, and using this periodicity, one can perform a Fourier expansion of the metric fields and the reduced phase space can be
decomposed into homogeneous $(\Gamma_{\rm{hom}})$ and inhomogeneous sectors $(\Gamma_{\rm{inhom}})$. The homogeneous sector is equivalent to the phase space of the compact vacuum Bianchi-I spacetime. For the inhomogeneous sector one can introduce the creation and annihilation variables, $a_m$ and $a_m^*$ respectively,
using which
%by treating the non-zero modes of the Fourier expansion of the metric  field, as corresponding to a massless scalar. Using these creation and
%annihilation operators,
the global diffeomorphism constraint can be written as
\be
C_\theta = \sum_{n = 1}^\infty \, n (a_n^\dagger a_n - a^*_{-n} a_{-n}) = 0 ~.
\ee
The Hamiltonian for the inhomogeneous modes consists of a free part $H_o$:
\be
H_o = \f{1}{8\pi G \gamma^2}\,\sum_{n \neq 0} \, |n| \, a_n^\star
a_n
\ee
and an interaction part $H_{\rm int}$:
\be
H_{\mathrm{int}} = \f{1}{8\pi G
\gamma^2}\, \sum_{n \neq 0} \f{1}{2 |n|} \, (2 a_n^\star a_n + a_n
a_{-n} + a_n^\star a^\star_{-n})
\ee
which mixes different inhomogeneous modes. The total Hamiltonian constraint for all the modes is given by
%non-zero modes can be expressed in terms of these creation and annihilation operators:
%$H^\xi_f = \sum_{n \neq 0} \, |n| \, a_n^* a_n$. Further, there is also an  interaction term,
%\be
%H_{\mathrm{int}}^\xi = \sum_{n \neq 0} \f{1}{2 |n|} \, (2 a_n^* a_n + a_n a_{-n} + a_n^* a^*_{-n}) ~
%\ee
%and together they form the non-trivial coupling between homogeneous and inhomogeneous sectors in the Hamiltonian constraint:
\ba
C_H &=& \nonumber - \f{2}{\gamma^2 V} \bigg[\left(c_\theta p_\theta \, c_\sigma p_\sigma + c_\theta p_\theta \, c_\delta  p_\delta + c_\sigma p_\sigma
\, c_\sigma p_\sigma \right) \, \\ && ~~~~~~~~~~~~~~~~~ - \, G \left(32 \pi^2 \gamma^2 |p_\theta| H_o + \f{(c_\sigma p_\sigma + c_\delta
p_\delta)^2}{|p_\theta|} H_{\mathrm{int}}\right) \bigg] ~,
\ea
where the  terms in the first parenthesis arise from the homogeneous part corresponding to the phase space of the Bianchi-I spacetime. This part of the
Hamiltonian constraint is loop quantized following the technique elaborated in Sec. \ref{secBianchi}, but for the earlier quantization \cite{chiou_bianchi}. The inhomogeneous part of the constraint, expressed in terms of the annihilation and creation operators, is Fock quantized. The resulting quantum constraint operator is
\vskip-0.8cm
\be \label{ghc3} \Theta_{\rm G} = \Theta_{\rm{B-I}}\, +\, \f{1}{8 \pi
\gamma^2} \Big[32 \pi^2 \gamma^2 \widehat{|p_\theta|} \hat H_o +
\left(\widehat{\f{1}{|p_\theta|^{1/4}}}\right)^2 (\hat \Theta_\sigma
+ \hat \Theta_\delta)^2
\left(\widehat{\f{1}{|p_\theta|^{1/4}}}\right)^2 \, \hat
H_{\mathrm{int}} \Big]\, \ee
where $\Theta_{\rm{B-I}}$ denotes the quantum Hamiltonian constraint operator for the Bianchi-I spacetime and the inverse powers of $\hat p_\theta$ are computed by
 expressing them in terms of a Poisson bracket between the positive powers of $p_\theta$ and the holonomies in the classical theory, and promoting the latter to a commutator \cite{tt}. Once we have these quantum constraints,
the inner product is obtained by using the same strategy as in the quantization of isotropic and anisotropic spacetimes. A complete set of Dirac observables are found which are self-adjoint with respect to the inner product. A feature of the above quantization, shared with the loop quantization of the Bianchi-I model, is the zero volume states in the homogeneous sector are decoupled by the action of
the Hamiltonian constraint. Thus the  states corresponding to the classical singularity at zero volume are absent and in this sense the
singularity is resolved.

The physics of the the Planck regime in these spacetimes is being explored using effective dynamics which confirms the existence of the bounce in the presence of
inhomogeneities. Earlier work in the
effective dynamics of the Gowdy model was based on using the effective Hamiltonian constraint for a slightly different quantization of the Bianchi-I spacetime \cite{chiou_bianchi}, which has some undesirable features \cite{awe2,cs3,ps11}.  It was found that in the case where the bounce can be approximated by the dynamics of the vacuum Bianchi-I model, the statistical average of the inhomogeneities across the bounce is positive. On the other hand, when dynamics is dominated by inhomogeneities the statistical
average of inhomogeneities across the bounce is preserved. More recently, the effective dynamics has been studied using the effective Hamitlonian constraint for the improved Bianchi-I quantization \cite{awe2}. It turns out that in comparison to the  homogeneous Bianchi-I spacetime in LQC, the inhomogeneities increase the volume at which the bounce occurs. 
  These investigations provide a
first glimpse of the bounce in the presence of Fock quantized inhomogeneities in LQC. It serves as a useful intermediate step towards the full loop
quantization of Gowdy spacetimes.

\section{Inhomogeneous perturbations in LQC}\label{sec4}

The  standard model of cosmology (see e.g.\ \cite{ll-book,sd-book,vm-book,sw-book}) is based on classical general relativity, and therefore cannot describe the earliest epochs of cosmic expansion, when curvature invariants reach the Planck scale. The goal of this and next section is to use LQC to extend existing  models to the Planck era. This extension opens the possibility of connecting Planck scale physics with cosmological observations, providing a new avenue to test some of the fundamental ideas on which the theory rests. The extension will also provide new physical mechanisms to account for the intriguing large scale anomalies observed in the CMB. 
 
But to carry out this task, the theoretical framework summarized in previous sections is insufficient. For, a  key ingredient in existing theories of the early universe, such as inflation, is the physics of  first order cosmological \emph{perturbations}, the so called scalar and tensor modes, that propagate on a classical Friedmann-Lema\^itre-Robertson-Walker  (FLRW) spacetime. We can observe features of these perturbations in the cosmic microwave background (CMB), allowing us to confront different models with observations. Therefore, if our goal is to incorporate LQC ideas in the description of the early universe, we need first to extend our theoretical framework to incorporate  inhomogeneous cosmological perturbations propagating on the \emph{quantum} cosmological spacetime, described in previous section. Since cosmological observations \cite{planck} have confirmed that the early universe is very well described by a homogenous an isotropic  FLRW line element with spatial curvature compatible with zero, the rest of this section will focus on quantum spacetimes with these features, that were described in section \ref{sec2}.
 
Incorporation of inhomogeneous perturbations on quantum spacetimes poses an interesting challenge. Far from the Planck regime, when the quantum aspects of gravity can be neglected,  inhomogeneous first order perturbations are accurately described as quantum fields propagating in a classical expanding universe. The theory of quantum fields in curved spacetimes \cite{birrelldavies82,waldbook,parker-book}, well established since 1970's, provides the suitable theoretical arena. But in the Planck regime the background geometry is fully quantum. Since quantum states $\Psi(\nu,\phi)$ introduced in previous sections provide only probabilistic amplitudes for the occurrence of various metrics, a priori we do not have a classical FLRW geometry in the background.
How do inhomogeneous perturbations propagate on theses quantum geometries $\Psi(\nu,\phi)$? This section will show that the answer to this question becomes tractable under the  assumption that first order perturbations produce negligible back-reaction on the background quantum spacetime on which they propagate, i.e. when they can be considered as {\em test fields}. This assumption plays also a key role in the standard cosmological model, as well as in alternatives to inflation \cite{cbz2011,kost2002}. In LQC, once the test field approximation is made, it becomes possible to obtain a well-defined {\em quantum field theory of cosmological perturbations on  quantum FLRW spacetimes} \cite{akl}. The reader is referred to \cite{akl,aan2,puchta,dapor, dapor1,dapor2} for further details. (See also Refs. \cite{gambinipullinbh} for application of similar techniques to spherically symmetric spacetimes.) 

We begin by briefly summarizing the well-known classical theory of cosmological, gauge invariant, first order perturbations on spatially flat FLRW spacetimes (see e.g.\ \cite{reportbrandenberger} for details). This will establish notation and provide the arena for quantization. Then we construct the quantum theory upon it. Finally, we will use these results to include Planck scale physics in the description of the early universe, and to describe mechanisms to connect Planck scale physics with cosmological observations.

\subsection{Cosmological perturbations in classical FLRW spacetimes}
\label{sec:3.a}

Most explorations of the early universe rest on the assumption that the energy-momentum budget during the first stages of cosmic expansion was dominated by a scalar field $\phi$  commonly  called the inflaton, that is subject to an effective potential $\v(\phi)$. (An example was discussed in section 3.1.3, with $\v(\phi) = \frac{1}{2}m^{2}\phi^{2}$.) This scalar field can be thought either as a fundament or an effective degree of freedom. Then, motivated by CMB  observations which show that the early universe was extraordinarily homogenous and isotropic, one looks for solutions of Einstein equations given by a FLRW metric $g_{ab}(t)$ with a homogeneous and isotropic scalar field $\phi(t)$ as source, {\em together with inhomogeneous first order perturbations} $\delta g_{ab}(\vec{x},t)$, $\delta \phi(\vec{x},t)$. The gravitational field by itself contains two physical degrees of freedom---so most of the metric components are purely coordinate (gauge) dependent functions---and adding the scalar field our system has three physical degrees of freedom in total. In order to avoid gauge artifacts, it is convenient to re-write the perturbation fields  $\delta g_{ab}(\vec{x},t)$, $\delta \phi(\vec{x},t)$ in terms of these gauge invariant degrees of freedoms: they are made of the so-called scalar perturbation $\Q(x)$ ---known as the Mukhanov-Sasaki variable---and two tensor perturbations $\T^{(1)}(x)$ and $\T^{(2)}(x)$. Physically one can think of $\Q$ as representing perturbations of the scalar field, and tensor modes as representing the two polarization of a gravity wave. These variables, together with their conjugate momenta and the homogeneous and isotropic degrees of freedom, span the physical phase space $\Gamma_{\rm phys}=\Gamma_{\rm hom}\times \Gamma_{\rm pert}$, where $\Gamma_{\rm hom}$ is  made of two pairs of canonically conjugate variables $(a,\pi_{(a)};\phi,p_{(\phi)})$, with $a$  the standard scale factor of the FLRW metric and $\Gamma_{\rm pert}$ is the phase space of scalar and tensor modes. Since the two tensor modes behave identically, from now on we will denote them collectively by $\T$. 

We now discuss dynamics in  $\Gamma_{\rm phys}$. First, if we restrict to the homogenous sector, $\Gamma_{\rm{hom}}$, dynamical trajectories are generated by the restriction to FLRW of the Hamiltonian constraint of general relativity (see section \ref{sec2})
\be 	\mathcal{C}_{H}[N] =N\left[-\frac{3V_0}{8\pi G} \frac{p_{(a)}^2}{a}+\frac{1}{2} \frac{p_{(\phi)}^2}{a^{3}V_0}+a^3 V_0 \, \v(\phi) \right]\, ,\ee
where $\kappa=8\pi G$. The lapse function $N$ indicates the time coordinate one is using: $N=1$ corresponds to standard cosmic or proper time $t$, $N=a$ to conformal time $\eta$, and $N=V_0^3a^3/p_{(\phi)}:=N_{\phi}$ to choosing the scalar field $\phi$ as a time variables, which turns out to be the most appropriate choice in the LQC.
\footnote{As mentioned in section 3.1.3, in general we only have a `local clock' since $\phi$ is only a good time variable in patches of dynamical trajectories along which $\phi$ is monotonic.} 
The evolution generated by $\mathcal{C}_{H}[N]$ takes place entirely in $\Gamma_{\rm hom}$; it does not involve inhomogeneous perturbations. 

Dynamics in $\Gamma_{\rm pert}$ is generated by a true Hamiltonian $\mathcal{C}_2$, which is obtained from the second order piece of the scalar constraint of general relativity by keeping only terms which are quadratic in first order perturbations. This Hamiltonian has the form $\mathcal{C}_{2}=\mathcal{C}^{(\Q)}_{2}+\mathcal{C}^{(\T^{(1)})}_{2}+\mathcal{C}^{(\T^{(2)})}_{2}$, where in Fourier space
\be \label{pert-ham} \mathcal{C}_2^{(\T)}[N]= \f{N}{2 (2\pi)^3 }\,\, \int d^3 k \, \left( \f{4 \kappa}{a^{3}}\, |\pp^{(\T)}_{\vk}|^2 + \frac{a\, k^2}{4 \kappa} |\T_{\vk}|^2 \right)\, , \ee
\be \label{pert-hams} \mathcal{C}_2^{(\Q)}[N]= \f{N}{2 (2\pi)^3 }\,\, \int d^3 k  \, \left( \f{1}{a^{3}}\, |\pp^{(\Q)}_{\vk}|^2 + a\, (k^2+\mathcal{U}) |\Q_{\vk}|^2 \right) \, . \ee
Here $\mathcal{U}= [\v(\phi) \, r-2 \v_{\phi}(\phi) \sqrt{r} + \v_{\phi\phi}(\phi) ]a^2$, with $r=(3\kappa  \pphi^2/((1/2)\pphi^2+V_0^2 a^6 \v(\phi))$, and $\v_{\phi}(\phi)$  and $ \v_{\phi \phi}(\phi)$,  the first and second derivatives of the inflaton potential $\v(\phi)$ with respect to $\phi$. Scalar and tensor modes evolve independently of each other.

The equations of motion for tensor and scalar perturbations generated by the Hamiltonian (\ref{pert-ham}) and (\ref{pert-hams}) take the form, in conformal time $\eta$ %
\be  \label{waveeqs}
{\T}_{\vk}^{\prime\prime} + 2
\f{{a}^\prime}{a}\, {\T}_{\vk}^\prime + k^2
{\T}_{\vk} = 0 \, ; \hspace{1cm} {\Q}_{\vk}^{\prime\prime} + 2 \f{{a}^\prime}{a}\, {\Q}_{\vk}^\prime + (k^2 +{\mathcal{U}}(\eta)) {\Q}_{\vk} = 0 ~.\,\ee
In physical space this equations are
\be \Box \T(x)=0\hspace{1cm}  \Box \Q(x)-\mathcal{U}\,  \Q(x)=0\, , \ee
where $\Box=\nabla_a\nabla^a$ is the D'Alembertian of $g_{ab}$. Therefore, tensor modes satisfy the same equation as a massless scalar  field in FLRW, and similarly for scalar perturbations, except for the presence of the external potential $\mathcal{U}$. As expected,  the dynamics in $\Gamma_{\rm pert}$ knows about $\Gamma_{\rm hom}$: scalar and tensor perturbations satisfy linear differential equations (\ref{waveeqs}) which contain coefficients involving background variables. Dynamics then is obtained by {\em first} solving the evolution for $a(\eta)$ and $\phi(\eta)$ using only $\mathcal{C}_{H}$, and {\em then} `lifting'  the resulting dynamical trajectory to $\Gamma_{\rm pert}$ using (\ref{waveeqs}). This is a consequence of the main approximation underlying this construction: perturbations produce negligible back-reaction on the background metric. 

In the inflationary scenario, the next step is to quantize the perturbation fields $\Q$ and $\T$, but keeping the homogenous degrees of freedom as classical. This is justified because curvature invariants are well below the Planck scale at all times during and after inflation, and hence quantum effects of the gravitational background are expected to be negligible. Therefore, one keeps the homogenous phase space  $\Gamma_{\rm hom}$ unmodified, but replaces the classical phase space of perturbations $\Gamma_{\rm pert}$ by a Hilbert space $\mathcal{H}_{\rm pert}$ in which the perturbation fields $\hat \Q$ and $\hat \T$ are represented as quantum operators. This is a quantum filed theory in a classical FLRW spacetime, and therefore well established techniques \cite{ll-book,sd-book,vm-book,sw-book} are available to extract physical predictions from this system. 

As in the classical theory, this semiclassical framework rests on the test field approximation. A necessary condition for its validity is that  the stress-energy of perturbations must be subdominant compared to the background contribution. Since energy and momentum are quadratic in the basic variables, one needs to introduce renormalization techniques to obtain well-defined expressions. The ambiguities in the process of renormalization in curved spacetimes add difficulties in testing the validity of this semiclassical theory. 

In the next subsection we describe the framework in which  {\em both} the homogenous as well as the inhomogeneous degrees of freedom are treated quantum-mechanically.

\subsection{Quantum theory of cosmological perturbations on a quantum FLRW \label{QFTQST}}

The description of perturbations in a quantum cosmological background is more complicated that in the classical FLRW case, although it
will follow  the same logical steps. In particular, the constructions relies upon the assumption that perturbations produce negligible effects on the background. In the classical theory, the absence of back-reaction is reflected in the fact that dynamics of background fields $a(\eta)$ and $\phi(\eta)$ is completely independent of perturbations. In the quantum theory, the test field approximation is  incorporated by assuming that the total wave function $\Psi$ has the form of a product 
\be \label{tenpro}  \Psi(a, \phi,\Q_{\vk}, \T_{\vk})=\Psi_{\rm hom}(a,\phi)\otimes \Psi_{\rm pert}(a, \phi,\Q_{\vk}, \T_{\vk})\, . \ee
This structure implies the absence of correlations between background and inhomogeneous degrees of freedom initially, which is then maintained during evolution as long as the test field approximation holds. Our task now is, first, to construct the quantum theory describing $\Psi_{\rm hom}$,  and then study the evolution of  $\Psi_{\rm pert}$ on the background geometry $\Psi_{\rm hom}$.

The evolution of $\Psi_{\rm hom}$ follows exactly the steps described in section 2 except that, as noted in section 3.1.3, in presence of an inflationary potential $\v(\phi)$, the  operator $\sqrt{\Theta}$ appearing in the evolution equation (\ref{qhc5})  now must be replaced by\\ $\big{|}\Theta - \v(\phi) \, \nu^2 \, \frac{\pi G\gamma^2}{2}\big{|}^{1/2}$. The presence of $\v(\phi)$ poses new challenges, because the resulting operator  fails to be essentially self-adjoint for a generic potential $\v(\phi)$.  However, as we will see in the next subsection, we will be only interested in situation in which $\langle \Theta \rangle \gg \langle \v(\phi) \, \nu^2 \, \frac{\pi G\gamma^2}{2} \rangle$ in the Planck era. Under these circumstances, the potential can be treated as a perturbation to $\Theta$, and one can show it produces a negligible contribution to the evolution in the quantum gravity regime. More precisely, as discussed in \cite{aag}, in situations of physical interest (see next subsection), the evolution with and without potential produce results for observable quantities with differences several orders of magnitude smaller than observational error bars.  One can therefore obtain reliable predictions without including the potential in the quantum gravity regime. Then, we can directly import the result of section \ref{sec2} for the evolution of $\Psi_{\rm hom}$. Nevertheless  the mathematical subtleties appearing in the inclusion of the inflaton potential constitute an important open issue, although not of direct relevance for the phenomenological considerations. These issues have been studied for a constant potential, i.e.\ a cosmological constant $\Lambda$, in \cite{ap} where it  was found that, although the operator  $\big{|}\Theta - \Lambda \,  \nu^2 \, \frac{\pi G\gamma^2}{2}\big{|}^{1/2}$  fails to be essentially self-adjoint, the quantum evolution is surprisingly insensitive to the choice of the self-adjoint extension.

The next task is to construct the theory of perturbations $\Psi_{\rm pert}$ propagating on the quantum spacetime  $\Psi_{\rm hom}$. Dynamics in this theory is extracted from the constraint equation (\ref{qhc5}) in presence of perturbations. For briefly, we will write down some of the intermediate steps of the quantization for tensor modes, and simply provide the result for scalar perturbations at the end. 

Evolution for the entire system will be obtained from the  constraint equation (\ref{qhc5}), that now reads 
 \be -i\hbar\, \partial_\phi (\Psi_{\rm hom}\otimes \Psi_{\rm pert}) =  \big{|}\hbar^2 \Theta - 2V_0\,  {\mathcal{C}}^{(\T)}_{2}\big{|}^{\frac{1}{2}} \, (\Psi_{\rm hom}\otimes \Psi_{\rm pert})\, \ee
The test field approximation allows us to use perturbation theory to solve this equation. We will treat $\Theta$ as the Hamiltonian of the  `heavy' degree of freedom and $ {\mathcal{C}}^{(\T)}_{2}$ as the Hamiltonian of the light one. Then, the previous equation can be approximated by (see \cite{akl} for details)
\ba  (-i\hbar\, \partial_\phi \Psi_{\rm hom})\otimes \Psi_{\rm pert}&+&\Psi_{\rm hom}\otimes  (-i\hbar\, \partial_\phi \Psi_{\rm pert})=\nonumber \\
&=& \hbar \, \sqrt{\Theta} \Psi_{\rm hom} \otimes \Psi_{\rm pert}- {\mathcal{C}}^{(\T)}_{2}[N_\phi] (\Psi_{\rm hom}\otimes \Psi_{\rm pert})\, \ea
where we have used that, while the $\Theta$ operator  acts   on $\Psi_{\rm hom}$ but not on $\Psi_{\rm pert}$,   $ {\mathcal{C}}^{(\T)}_{2}[N_\phi]$ in contrast acts on both states, since it contains  background as well as perturbation operators. 
The first term in each side of the previous equality cancel out by virtue of the evolution of the background state (\ref{qhc5}), and we are left with 
\be \Psi_{\rm hom}\otimes  (i\hbar\, \partial_\phi \Psi_{\rm pert})= {\mathcal{C}}^{(\T)}_{2}[N_\phi] (\Psi_{\rm hom}\otimes \Psi_{\rm pert})\, \ee
This equation tells us that in the test field approximation the right hand side is proportional to $\Psi_{\rm hom}$, and we therefore can take the inner product of this equation with $\Psi_{\rm hom}$ without losing  information. The last equation then reduces to 
\be  \label{qftqst} i\hbar\, \partial_\phi \Psi_{\rm pert}=\langle  {\mathcal{C}}^{(\T)}_{2}[N_\phi] \rangle\,  \Psi_{\rm pert}\, , \ee
where the expectation value is taken in the physical Hilbert space of the homogeneous sector. In other words, as long as the test field approximation holds, the evolution of perturbation is obtained from ${ \mathcal{C}}^{(\T)}_{2}[N_\phi] $ where the \emph{background operators} are replaced by their expectation value in the state $\Psi_{\rm hom}$.

This theory is conceptually different from quantum field theory (QFT) in classical spacetimes. As previously mentioned , the spacetime geometry is not described by a  classical metric,  but it is rather characterized by a wave function $\Psi_{\rm hom}$ that contains the  quantum fluctuations of the geometry. Equation (\ref{qftqst}) tell us that, indeed, the propagation of perturbations {\it is} sensitive to those fluctuations, and not only to the mean-value trajectory of the scale factor $\langle \h a \rangle$. 

An interesting aspect of the previous evolution equation (\ref{qftqst}) is that, by simple manipulation, it can be written as 
\be \label{Teqn} \hat{\T}_{\vk}^{\prime\prime} + 2
\f{\tilde{a}^\prime}{\tilde{a}}\, \hat{\T}_{\vk}^\prime + k^2
\hat{\T}_{\vk} = 0 \, ,\ee
where we have defined
\be \label{at}  \quad\quad \tilde{a}^4 := \f{\langle
\hat{\Theta}^{-\f{1}{4}}\, \hat{a}^4(\phi)\,
\hat{\Theta}^{-\f{1}{4}}\rangle}{\langle \hat{\Theta}^{-\f{1}{2}}\rangle}\, ,
\ee
and the quantum conformal time $\tilde \eta$ is related to the internal time by 
\be \label{etat} \d\tilde{\eta} := \tilde{a}^2(\phi)\,
\langle
\hat{\Theta}^{-\f{1}{2}}\rangle \, \d\phi\, . \ee
But interestingly,  (\ref{Teqn})  has the same form as the equation of the field $\hat{\T}$ propagating on a smooth FLRW geometry (see eqn. (\ref{waveeqs})). More explicitly,  the evolution (\ref{Teqn}) is {\em mathematically indistinguishable} from a QFT on a {\em smooth FLRW metric} $\tilde g_{ab}$
\be \tilde{g}_{ab}\, \d x^a \d x^b :=\tilde{a}^2(\tilde\eta)\, (-\d\tilde\eta^2 +
\, \d\vec{x}^2) \ee
In position space, the equation for $\hat \T$ reads\,\, $\Box \hat \T(x)=0$,\, where $\Box$ is the d'Alembertian of the metric $\tilde g_{ab}$. In a similar way the scalar perturbations satisfy the second order differential equation
\be \label{Qeqn} \hat{\Q}_{\vk}^{\prime\prime} + 2 \f{\tilde{a}^\prime}{\tilde{a}}\, \hat{\Q}_{\vk}^\prime + (k^2 +\tilde{\mathcal{U}}(\tilde\eta)) \hat{\Q}_{\vk} = 0 \, .\ee
Scalar perturbations propagate in the {\em same}  metric $\tilde g_{ab}$ as tensor modes but, additionally, they feel a  potential given by
\be \label{ut} \tilde{\mathcal{U}}= \f{\langle \hat{\Theta}^{-\f{1}{4}}\, \hat{a}^2 \, \hat{\mathcal{U}}\, \hat{a}^2  \hat{\Theta}^{-\f{1}{4}}\rangle}{\langle \hat{\Theta}^{-\f{1}{4}}\, \hat{a}^4 \, \hat{\Theta}^{-\f{1}{4}}\rangle} \, ,\ee
where $\h {\cal{U}}$ is the operator associated to the classical external potential $\cal{U}$ written after eq. (\ref{pert-hams}).\\

Therefore, at the practical level, in order to evolve test fields on a quantum geometry $\Psi_{\rm hom}$ one only needs to compute the  components of $\tilde g_{ab}$ from $\Psi_{\rm hom}$, and them proceed as in standard quantum field theory in curved spacetimes. 
 
The following remarks are in order:

(i) No further assumptions beyond the test field  approximation have been made to obtain this result. In particular, the state $\Psi_{\rm hom}$ is not assumed to have small quantum dispersion in the configuration or momentum variables \cite{aag}. 

(ii)  The metric $\tilde g_{ab}$ does not satisfy Einstein's equations. This is obvious from its definition: its coefficients are obtained as expectation values of background operators in the homogenous and isotropic quantum geometry $\Psi_{\rm hom}$, and therefore they depend on $\hbar$. Rather, $\tilde g_{ab}$ is a mathematical object that neatly captures the information in $\Psi_{\rm hom}$ that is relevant for the propagation of tensor and scalar modes. It is remarkable  that, from the rich information contained in $\Psi_{\rm hom}$,  test fields only `feel' a few of its moments, namely (\ref{at})(\ref{etat})(\ref{ut}), and furthermore, that these moments can be codified in a smooth metric $\tilde g_{ab}$! This metric is called in the literature the {\em effective dressed metric}.\cite{akl}. Effective because it contains all the information in $\Psi_{\rm hom}$ that is relevant for perturbations; and dressed because it depends not only  on the mean value of  $\Psi_{\rm hom}$ but also on some of its quantum 
fluctuations. 

(iii) One should not think about  $\tilde g_{ab}$ as approximating  the physical background geometry in any way. The spacetime geometry  $\Psi_{\rm hom}$ is quantum, and in general cannot be approximated in any reasonable sense by a smooth metric tensor. Rather, $\tilde g_{ab}$   encodes only the information in $\Psi_{\rm hom}$ \emph{that test fields care about.} In other words, if we probe the quantum geometry $\Psi_{\rm hom}$ using \emph{only} tensor and scalar modes, we will be unable to distinguish it from  a smooth geometry characterized by $\tilde g_{ab}$. But  if we  use other observables are used, e.g.\  powers of  curvature invariants,  we would easy realized that the background gravitational field has additional information not captured by $\tilde g_{ab}$. 
 
(iv) If the state $\Psi_{\rm hom}$ is chosen to have very small quantum dispersion in the volume $\nu$ (equivalently in the scale factor $a$), then  the dressed metric $\tilde g_{ab}$ is indistinguishable from the effective metric discussed in section (\ref{sec.eff}). 
  
Now, because the theory of tensor and scalar test fields propagating in $\Psi_{\rm hom}$ has been written as a QFT in curved spacetime $\tilde g_{ab}$,  we can import the well-known theoretical machinery developed in that context and construct a Fock-type quantization of the test fields. Then, we end up with a hybrid quantization approach, following the ideas introduced in Ref.\ \cite{hybrid1} to study Gowdy models, in which homogenous degrees of freedom are quantized  using LQG techniques and inhomogeneous fields using standard Fock techniques. This strategy is mathematically consistent and physically attractive, and it allows to make contact with the treatment of test fields in standard cosmology, e.g.\ in the inflationary scenario. 

 As in the semi-classical theory, testing the validity of the test field approximation is not a simple task. In Ref. \cite{aan2,aan3},  this has been done following the same steps as in the semi-classical theory, namely by comparing the expectation value of the renormalized energy and pressure of perturbations with the background contributions. The test field approximation may be a real limitation in situations of physical interest, and current efforts are focused on extending the range of applicability of framework summarized  above \cite{agm}.

\subsection{Other approaches}

The LQC literature is vast, and other approaches to the quantization of first  order perturbations exist (see  \cite{calcagni, lqcreview, bbcg, barraureview, reviewmena, madrid1, madrid2, wilson-ewing2,we2016, grainreview2016}). The `hybrid quantization approach' \cite{madrid2, madrid3, madrid4, madrid1, madrid5,benitezolmedo2016,mdbo2016}, originally suggested in Ref. \cite{hybrid1}  for Gowdy cosmologies, has  similarities with the `dressed metric approach' presented above, particularly the fact that test fields  are  quantized  using standard Fock  techniques, while the background FLRW geometry follows the non-perturbative methods of loop quantum gravity. The two methods however differ in the way in which constraints for perturbations are imposed, already at the classical level. But these conceptual difference  have little impact in observable quantities, at least in the set of solutions that have been explored so far. Hence, the predictions for the primordial spectrum of perturbations are very similar in both approaches, adding robustness to the program. We focus on the `dressed metric approach' on this chapter primarily because so far the most detailed calculations leading to phenomenological predictions have been carried out in this framework.

The `separate universe' approach in LQC \cite{wilson-ewing2,we2016} has the conceptual advantage of quantizing the homogenous and inhomogeneous degrees of freedom `in tandem', using only loop techniques, at the expenses of being only applicable to very long wavelengths. 

The `anomaly-free quantization approach' summarized in Refs. \cite{bbcg,bojowald-paly,grainreview2016,barraureview} uses the algebra of gravitational constraints as guiding principle, and develops an effective approach in which quantum gravity corrections are incorporated. The expressions of these quantum corrections are guided by imposing that the constraint algebra closes at the desired order in perturbations. This program produces a physical picture of the very early universe which is very different from those in other LQC approaches. Its phenomenological consequences have been explored in \cite{sbblmg2016,bbgs2016,bg2016}, where it is claimed that some of the predictions are in sharp disagreement with CMB observations. (For additional discussion, see \emph{ Chapter 8}.)

\section{Application: LQC extension of the inflationary scenario}\label{sec5}

Now that we have a quantization for FLRW spacetimes and a theory of scalar and tensor perturbations propagating thereon, we are ready to apply this theoretical framework to the early universe.  Among the existing models, the inflationary scenario is perhaps the most accepted one, and this section  summarizes an approach to  extend it to the quantum gravity  regime \cite{aan1,aan3}. See  \cite{cwe2014b,cqswe2014,cwe2014} for interesting work in LQC in the context of the ``matter bounce scenario''. %The goal of this section will be twofold: (i) to show how quantum gravity ideas can be useful to extend our understanding of the very early universe; and, (ii) to establish a promising avenue to connect quantum gravity ideas with observations. 

\subsection{The strategy}

In the inflationary scenario one starts by assuming that tensor and scalar modes are in the Bunch-Davies  vacuum at some early time when inflation begins. This state is then evolved until the end of inflation and relevant quantities for observations, e.g.\ the power spectra of tensor and scalar perturbations, are computed. The choice of the vacuum, however, rests on the implicit assumption that the evolution of the universe {\em before} inflation is unimportant in what the tensor and scalar modes respect; otherwise tensor and scalar modes would reach the onset of inflation in an excited state relative to the inflationary vacuum. But to show whether this is a reasonable assumption one needs a model for the pre-inflationary universe.  Our strategy is to use LQC for this purpose, and {\em compute}, rather than postulate, the state of inhomogeneous perturbations at the onset of the slow-roll phase. In the resulting picture the universe contracts for an infinite amount of time, bounces at the Planck scale, and then inflation starts at some time after the bounce\footnote{In concrete simulations inflation starts around $10^{-35} s$ after the bounce, and lasts for a similar interval of time. However, the universe only expands for around fifteen   e-folds from the bounce to the onset of inflation, while it expands for more than 60 e-folds during inflation. On the contrary, while the spacetime scalar curvature during inflation is almost constant, in decreases about eleven orders of magnitude from the bounce to the onset of slow-roll. These numbers are obtained using initial conditions that lead to interesting observable effects in the CMB, and they can vary for other choices.} The strategy will then be to start with vacuum initial conditions for tensor and scalar modes at early times, possibly prior to the bounce, evolve them using the LQC equations, and show that the resulting state at the onset of inflation coincides with the Bunch-Davies vacuum to a good approximation. 
If the initial state can be  specified in a compelling fashion and if it does not evolve to a state that is sufficiently close to the Bunch-Davies vacuum at the onset of inflation, the viability of the framework would be jeopardized. If, on the other hand, the evolved state turns out to be close to the Bunch-Davies vacuum but with appreciable deviations, there would be new observable effects. 

Therefore, the effect of LQC in observable quantities that we are looking for does {\em not} come from quantum gravity corrections generated {\em during inflation}. There, the energy density and curvature of the universe are around eleven orders of magnitude below the Planck scale, and  quantum gravity corrections are  suppressed by a similar factor. On the contrary, the effects we are looking for are generated {\em before} inflation, when quantum gravity effects dominate. The important fact is that scalar and tensor perturbations keep memory of these effects \cite{agulloparker,agullo-navarro-salas-parker} which then  can be imprinted in the CMB temperature anisotropies---if the amount of inflationary expansion is not much larger than 70 e-folds for these corrections not to be red-shifted to super-horizon scales.

The exploration of phenomenological consequences in LQC follows these steps:

(i) Chose an inflationary potential $\v(\phi)$. 

(ii) Specify the quantum state for the FLRW geometry $\Psi_{\hom}(\nu,\phi)$. 

(iii) Specify the quantum state for tensor and scalar modes $\Psi_{\rm pert}$. 

(iv) Evolve the background and perturbations using the theoretical framework spelled out in previous sections. 

(v) Compute observable quantities.\\

We now discuss these steps in more detail:\\

(i) Chose an inflationary potential $\v(\phi)$.  Although it would be desirable to derive the potential from first principles, at the present time there is no compelling candidate within LQC. One could expected though that  $\v(\phi)$ originates from a theory of particle physics, rather than from LQC which is a purely gravitational theory---although there also exist the exciting possibility that the inflaton field and its potential have a purely gravitational origin \cite{barrow1988,barrow1988b}. Therefore, the strategy so far in LQC has the same as in standard inflation, namely to use different phenomenologically viable potentials and contrast the results with observations. Two choices have been explored in great detail in LQC: the quadratic\footnote{Upper bounds on the tensor-to-scalar ratio recently obtained from the Planck satellite observations \cite{planck} slightly disfavor the quadratic potential. However, LQC corrections can alleviate these constraints \cite{am2015}, and therefore this potential may still be of some interest.} and the so-called Starobinsky potential \cite{gb2015, gb2015b} :
\be \v(\phi)=\frac{1}{2}m^2\phi^2 \quad {\rm and} \quad 
\v(\phi)=\frac{3 m^2}{32 \pi}\, (1-\exp(-\sqrt{16\pi G/3}\, \phi))^2 \ee
The free parameter $m$ can be fixed by CMB observations although, as analyzed in detail in \cite{am2015}, in LQC there is some extra freedom. It is also important to keep in mind that LQC effects have a purely quantum gravity origin and are largely independent of the choice of potential. Therefore, predictions obtained from different choices of $\v(\phi)$ are all quite similar. \\

(ii) Quantum state for the FLRW geometry $\Psi_{\hom}(\nu,\phi)$. 
As discussed at the beginning of section \ref{QFTQST}, the states $\Psi_{\hom}$ of interest differ from the ones described in section \ref{sec2} in that now the dynamics contains an inflationary potential $\v(\phi)$. But detailed analysis \cite{aan3,am2015,gb2015b} have showed that LQC effects can be  imprinted in the CMB only if the contribution from the potential $\v(\phi)$ is subdominant around the bounce time, as compared to the kinetic energy of $\phi$. This is because potential dominated bounces lead to very long inflationary phases, and  LQC effects will then be red-shifted to super-Hubble scales. But even if the potential is subdominant around the bounce time, it gains relevance later in the evolution, and eventually dominates during inflation. However, it turns out that the potential dominated regime occurs well after the quantum gravity era, at a time when the energy density and curvature invariants are all well below the Planck scale, and general relativity becomes an excellent approximation. Therefore, the 
regime of physical interest for  investigations of LQC phenomenology is such that the potential $\v(\phi)$ can be treated as a perturbation during the quantum gravity era. Furthermore, as shown in Ref. \cite{aag},  the relative effect of the potential in observable quantities under this circumstances is several order of magnitude below observational sensitivity. Therefore, one can choose to simply ignore the potential in the quantum gravity era without affecting the accuracy of observational predictions. In this situation one is led to work with the states described in section \ref{sec2}.

Among all the states for the homogeneous and isotropic  gravitational field described in section \ref{sec2}, the simplest choice is to work with states $\Psi_{\rm hom}(\nu,\phi)$ that have very small quantum dispersions in volume $\nu$ around the bounce time. These states remain `sharply peaked" during the entire evolution. More importantly, the resulting geometry can be accurately described by the effective metric presented in section (\ref{sec.eff}). For these states the energy density at the time of the bounce saturates  the supremum in the entire physical Hilbert space, $\rho_{\rm max}$. The only freedom in $\Psi_{\rm hom}(\nu,\phi)$ is then the way this energy density is divided between potential  and kinetic energy of the inflation field at the bounce. Different choices produce solutions that accumulate different amount of expansion between the bounce and the end of inflation, $N_{\rm B}:=\ln \big({a_{\rm end}}/{a_{\rm bounce}}\big)$; therefore, the freedom can be parameterized by the value of $N_{\rm B}$.

One can also choose states $\Psi(\nu,\phi)$ which have large dispersion in  $\nu$, and therefore are not sharply peaked in any of the variables. The computation of the scalar power spectrum is numerically more challenging in this case, and has been recently performed  in \cite{aag} for states containing relative quantum dispersion in $\nu$  as large as $168\%$ in the Planck epoch. The main lessons from this analysis are: i) observational quantities {\em are} sensitive to the quantum dispersion in  $\Psi_{\rm hom}(\nu,\phi)$; ii) however, the effects  are quite simple and, within observational error bars, predicted observable quantities cannot distinguish between a widely spread state which produces a certain amount of expansion $N_{\rm B}$, and a sharply peaked states with a slightly different valued of $N_{\rm B}$ (see \cite{aag} for details). Therefore, if one is only interested in observational predictions, there is no loss of generality in restricting to sharply peaked states, as long as different value of $N_{\rm B}$ are considered. This will be the strategy that we will follow in the rest of this section. \\

(iii) Quantum state for tensor and scalar modes. The specification of the initial state for inhomogeneous perturbations is an important question. Two main strategies have been followed in the LQC-literature: the state is specified by choosing vacuum initial conditions at \cite{aan1,aan3,am2015,mdbo2016} or before the bounce \cite{sbblmg2016,bbgs2016,am2015}. It could seems at first that the far past is the natural place to set up initial data for perturbations. One cannot disregard, however, the possibility that quantum gravity effects make the pre-bounce evolution unimportant. Note that in presence of an inflationary phase the radius of the observable universe is of the order of 10 Planck lengths at the time of the bounce. At these scales quantum gravity effects are very efficient, and there are indication that they could produce a diluting effect that makes scalar and tensor modes to forget about features acquired in the contracting phase. This is an interesting possibility, and in the absence of conclusive arguments the best strategy is to keep working with the two options and  contrast  their  predictions with observations. 

But even after making a choice between these two possibilities, one still has to face the inherent ambiguity in the definition of vacuum in quantum field theory in an expanding universe. If all wavelengths of interest are much smaller that the curvature radius, the adiabatic approach provides a useful criteria to reduce the ambiguity; if we are not in this situation, other arguments are needed. Different proposals for initial vacuum have been used in LQC. In \cite{ana}, in a more general context, a covariant criterion was introduced to specify the notion of vacuum at a given time, by demanding that the expectation value of the adiabatically renormalized  energy-momentum tensor vanishes at that time. In many situations of practical interest this condition provides a unique state. Ref.\ \cite{mdbo2016}, on the other hand, has fixed the freedom in the vacuum by demanding that certain time variations of the mode functions characterizing the quantum state in Fourier space are minimized during the evolution. Finally, in Refs.\ \cite{ag2016,ag2016b}  a quantum generalization of the Penrose's Weyl curvature hypothesis has been used to single out a vacuum state using considerations that bridge the Planck regime around the bounce and the end of inflation.\\

(iv) Evolution of the background and perturbations can be obtained by using the theoretical framework spelled out in previous sections.\\

(v) Computation of observable quantities. The  quantities of interest are the power spectrum of tensor and scalar perturbations, and their spectral indices, which are defined in the standard way (see e.g.\ \cite{sw-book}).\\

\subsection{Results}

\emph{Chapter 8} in this volume focuses on the phenomenology of loop quantum gravity. In particular, section 3 in that \emph{Chapter} is devoted to LQC, and includes results for CMB anisotropies obtained from the theoretical framework we have described in this chapter. Therefore, this section will be brief and its aim is to complement the analysis in the mentioned  chapter. 

\begin{figure}[tb] \begin{center}
  \includegraphics[width=4in]{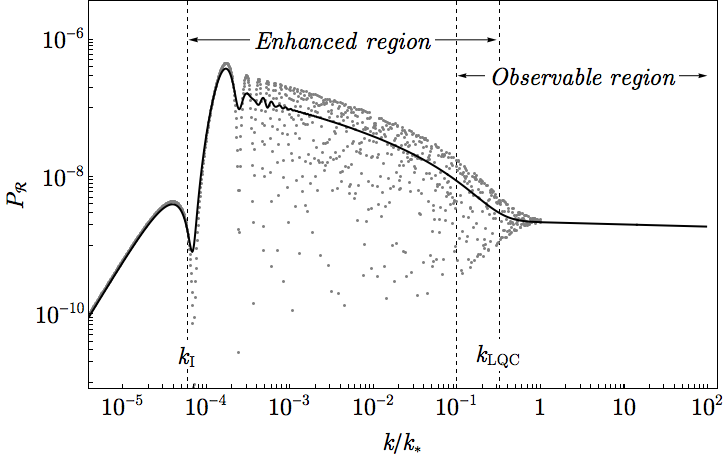} \\ \includegraphics[width=4in]{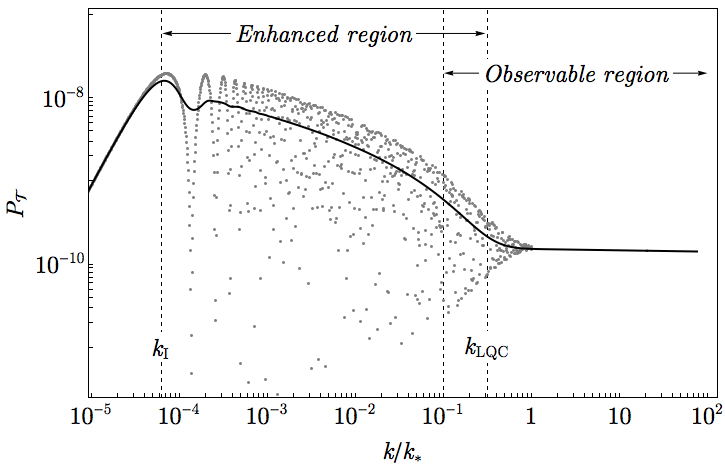}  \end{center}
    \caption{The LQC scalar (top) and tensor (bottom) power spectrum as a function of $k/k_*$, where $k_*$ is pivot  comoving scale corresponding to $0.002\, {\rm Mpc}^{-1}$ at the present time.  These plots are obtained for parameter values $m=1.3 \times 10^{-6}$,  preferred instantaneous vacuum \cite{ana} initial data for perturbations at initial time $t=-50000$, and value of the inflaton field at the bounce time $\phi_B=1$, all quantities in Planck units. The numerically evolved spectrum, shown in gray, is rapidly oscillatory; its average, shown in black, has an amplitude which is enhanced with respect to the standard predictions of slow-roll inflation for modes $k_I \lesssim k \lesssim k_{LQC}$ but agrees with them for $k\ll k_{\rm LQC}$.  The region of Fourier modes that are observable in the CMB for this choice of parameters is also shown; smaller $k$'s correspond to super-Hubble scales at the  present time. \label{spectrum}}
\end{figure}

Figure \ref{spectrum}, extracted from Ref.\ \cite{am2015}, shows the scalar and tensor  power spectrum for a given choice of parameters and initial state. The detailed analysis of these results was presented in \cite{aan1,aan3}, and further analyzed in \cite{am2015}.  The main new feature in the power spectra is the appearance of a new scale, $k_{LQC}$, which is directly related to the value of the spacetime scalar curvature  at the time of the bounce, $k_{LQC}/a(t_B):=\sqrt{R(t_B)/6}$. LQC corrections appear for Fourier modes with $k\lesssim k_{\rm LQC}$. Modes with larger $k$ are essentially insensitive to the bounce; they reach the onset of inflation in the Bunch-Davies vacuum and, as a consequence, their power spectrum is almost scale invariant. 

The observable effects of LQC in the CMB therefore depend on the value that the new scale, which is of the order of the Planck scale at the time of  bounce,  has at {\em the  present time} $t_0$, i.e.\  the value of  $k_{LQC}/a(t_0)$, which in turns is dictated by the number of e-folds from the bounce to the end of inflation $N_B$ (the expansion accumulated from the end of inflation until the present time is fixed by  the standard model of cosmology). If $k_{LQC}/a(t_0)$ is   larger than  ${k_*}/{a(t_0)}:=0.002 \, {\rm Mpc}^{-1}$, the observed power spectrum is predicted to deviate significantly from scale invariance. Available data from the CMB temperature anisotropies constrain $k_{LQC}/a(t_0)$ to be of the same order or smaller than $k_*/a(t_0)$, so $k_{\rm LQC}$ is constrained to be smaller than $k_{*}$. On the other hand, if $k_{LQC}/a(t_0)$ is smaller than approximately $0.1 ({k_*}/{a(t_0)})$, then LQC corrections are swept to length scales larger than our observable universe. Therefore, if LQC correction were to appear in CMB, we must have $k_{LQC} \in[0.1 k_*,k_*]$. This is equivalent of saying 
that  the amount of expansion from the bounce to the present time is such that  a wavelength of size twice the Planck length at the bounce is red-shifted to approximately $3000 {\rm Mpc}$ at the present time. Although there are no mechanisms based of precise arguments to explain why such coincidence should happen, Ref. \cite{ag2016} has provided concrete physical principles that lead to this situation. Furthermore, observations have detected deviations from the standard featureless scale invariant spectrum for  the low $k$ region of the power spectrum, indicating that new physics may be needed to account for the observed anomalies \cite{Planck2015Isotropy}. Although the associated statistical significant of these anomalies is inconclusive,  it is quite tempting to think that the may be visible traces of new physics, as indeed emphasized in \cite{Planck2015Isotropy}. 

A natural question is then whether the LQC bounce preceding inflation provides a suitable mechanism to quantitatively account for the observed anomalies. The Planck team has paid particular attention to two anomalous features in the CMB \cite{Planck2015Isotropy}, namely: i) a dipolar asymmetry arising from fact that the averaged power spectrum is larger in a given hemisphere of the CMB than in the other, and; ii) a power suppression at large scale, corresponding to a deficit of correlations at angular multipoles $\ell\lesssim 20$ as compared to the predictions of a scale invariant spectrum.  We now briefly summarize  existing ideas related to these anomalies in LQC.

A primordial dipolar asymmetry requires {\em correlations} between different  wave-numbers $k$ in the power spectrum. Such correlations do not arise at leading order in models for which the background is homogenous, as the scenario discussed in the last two sections: the two-point function in Fourier space is diagonal.  This motivated the authors of Ref. \cite{a2015} to go beyond  leading order and discuss the  corrections the primordial spectrum acquires from the three-point function (i.e.\ corrections from non-Gaussiantiy). As first point out in \cite{halo-bias3}, non-Gaussian effects in the two-point function could indeed be responsible of the observed dipolar modulation in the CMB. In Ref. \cite{a2015} this idea was implemented in LQC. The non-Gaussianity that inflation generates as a consequence of the pre-inflationary LQC bounce were computed and its effect on the primordial power spectrum were obtained. The result is that there exist values of the free parameters---the value of the inflaton field at the bounce $\phi_B$ (or equivalently, $N_B$) and its mass $m$---that make the non-Gaussian  modulation of the power spectrum to induce a scale dependent dipolar modulation in the CMB that agrees with the observed anomaly. Furthermore, this mechanism also offers the possibility  to account for the power suppression, since a {\em monopolar} modulation  appears, in addition to the dipole, at large angular scales, which could reverse the enhancement of power shown in figure \ref{spectrum}.  The analysis in \cite{a2015} included the non-Gaussianity generated during inflation, but a contribution to the three-point function from the bounce is also expected. However, this contribution to non-Gaussianity is significantly more challenging to compute, even numerically, because of the absence of the slow-roll approximation normally used to simplify the computations in inflation. Work is in progress \cite{abv2016} to complete this computation with the goal of establishing that the non-Gaussian modulation in LQC is a viable mechanism to simultaneously account for the two observed anomalies. 

Other ideas have also recently appear to account for the power suppression at large scales \cite{mdbo2016,ag2016}. They are related to the choice of initial state for scalar perturbation at the time of the bounce mentioned above in this section. The statement in these works is that one can find physical criteria to select a preferred notion of ground state at the bounce which, when evolved until the end of inflation, produce a power spectrum which is suppressed  compared to the standard scale invariant result for low values of $k$.

\section{Discussion}\label{sec6}

LQC provides a remarkable example of successful quantization of the sector of classical GR spacetimes with symmetries observed at cosmological scales. It is based on a precise mathematical framework, supplemented with sophisticated state of the art numerical techniques. One starts by showing that the requirement of background independence is strong enough to uniquely fix the quantum representation, just as the Poincar\'e symmetry  symmetry fixes the representation of the observable algebra in the standard quantum theory of free fields. One then uses this preferred representation. This procedure was first applied to a spatially flat FLRW background and the resulting quantum geometry was analyzed in detail. As described in this chapter, the final picture realizes many of the intuition that physicists, starting from Wheeler, have had about non-perturbative quantum gravity. Furthermore, interesting questions can now be answered in a precise fashion in LQC. Of particular interest is the way in which quantum effects are able to overwhelm the gravitational attraction and resolve the big bang singularity. While the LQC non-perturbative corrections dominate the evolution in the Planck regime and remove the big bang singularity, they disappear at low energies restoring agreement with the classical description. This is a  non trivial result. The analysis has been extended to more complicated models containing spacial curvature, anisotropies, and even models with infinitely many degrees of freedom such as the Gowdy spacetime, adding significant robustness to the emergent physical picture. Using effective spacetime description of LQC, the problem of singularities in general has been addressed, which provides important insights on the generic resolution of strong curvature singularities.

One can further extend the regime of applicability of LQC by including cosmological perturbations. In standard cosmology one describes scalar and tensor curvature perturbations by quantum fields propagating in a classical FLRW spacetime. This is the theoretical framework ---QFT in classical spacetimes--- on which the phenomenological explorations of the early universe rely, e.g.\ in the inflationary scenario. In this chapter we have reviewed how such a framework can be generalized by replacing the classical spacetime by the quantum geometry provided by LQC. This framework provides a rich environment to analyze many interesting questions both conceptually and at the phenomenological level. It offers the theoretical arena to explore the evolution of scalar and tensor perturbations in the early universe, and to provide a self-consistent quantum gravity completion of the standard cosmological scenarios. It is our view that the level of detail and mathematical rigor attained in LQC is uncommon in quantum cosmology. The new framework has become a fertile arena to obtain new mechanisms that could explain some of the anomalous features observed in the CMB, which indicate that  physics beyond inflation is required to understand the large scale correlations in the CMB \cite{Planck2015Isotropy}.  

Since LQC is a quantization of classical spacetimes with symmetries that are appropriate to cosmology, the theoretical framework shares the limitations of the symmetry reduced quantization strategy. Symmetry reduction often entails a drastic simplification, and therefore one may loose important features of the theory by restricting the symmetry prior to quantization. This is an important issue which has attracted efforts from different fronts. First let us recall that the BKL conjecture further supports the idea that quantum cosmological models are very useful in capturing the dynamics of spacetime near the singularities. Within LQC itself, the concern was initially alleviated by checking that models with larger complexity, such as anisotropic Bianchi I model, correctly reproduced the  FLRW quantization previously obtained, when the anisotropies are `frozen' at the quantum level. This test is even more remarkable when applied to models that have infinitely many degrees of freedom to begin with, as it is the case of the Gowdy model. More generally, there are interesting recent results on establishing a connection between LQC and LQG \cite{engle,brunn,tim}. These include \emph{quantum-reduced} loop quantum gravity  \cite{alesci-cianfrani}, where the main idea is to capture symmetry reduction at the quantum level in LQG and then pass to the cosmological sector, and group field theory cosmology \cite{gftcosmo}. Promising results have been obtained in these approaches. As examples, improved dynamics as the one used in isotropic LQC has been found in quantum-reduced loop quantum gravity \cite{alesci-improved}, and evidence of LQC like evolution and bounce have been reported in group field theory cosmology \cite{ed-bounce}. It is rather encouraging that results from different directions seem to yield a consistent picture of the Planck scale physics as has been extensively found in LQC

Another important ingredient in LQC is the process of de-parameterization. In the absence of a fundamental time variable in quantum gravity, in LQC one follows a relational-time approach in which one of the dynamical variables plays the role of time, and one studies the evolution of other degrees of freedom with respect to it. As explained in section \ref{sec2}, in most of the LQC literature one uses a massless scalar field as time variable. An important question is how the physical results depend on the variable chosen as a time, i.e. if quantum theories constructed from different relational times are unitarily related. This is an age old question in quantum cosmology, but so far has not been systematically addressed.

It is worth commenting on some of the directions where significant progress has been made in LQC, in contrast to the earlier works in quantum cosmology. The first one deals with a rigorous treatment of fundamental questions in  quantum cosmology about the probability of events -- such as the probability for encountering a singularity or a bounce. These are hard questions whose answers had been elusive due to the lack of sufficient control over the physical Hilbert space structure, including properties of observables and a notion of time to define histories.  Thanks to the quantization of isotropic and homogeneous spacetimes using a scalar field $\phi$ as a clock, a consistent histories formulation can be completed both in the Wheeler-DeWitt theory and LQC \cite{consistent1,consistent2, consistent3,consistent4}. A covariant generalization of these results has also been pursued  \cite{consistent5}. Using exactly soluble model of sLQC computation of class operators, decoherence functional and probability amplitudes can be performed. It turns out that in the Wheeler-DeWitt theory the probability for bounce turns out to be zero even if one considers an arbitrary superposition of expanding and contracting states. The probability of bounce turns out to be unity in LQC. These developments show that not only LQC has been successful in overcoming problem of singularities which plague Wheeler-DeWitt theory, it has also established an analytical structure which has been used to answer foundational questions both in LQC and the Wheeler-DeWitt theory.

The second direction where developments in LQC are expected to have an impact beyond LQG are in the development of sophisticated numerical algorithms to understand the evolution in deep Planck regime for a wide variety of initial states, including with very large spreads \cite{dgs1,dgs2,squeezed}. Some of these techniques have been exported from traditional numerical relativity ideas which are modified and applied in the quantum geometric setting. Using high performance computing, these methods promise to yield a detailed picture of the 
physics of the Planck scale. These techniques can be replicated in a straightforward way for other quantum gravity approaches. More importantly they provide a platform to understand the structure of quantum spacetime analogous to the numerical works in classical gravity \cite{berger,dg1}. A deeper understanding of how quantum gravitational effects modify the BKL conjecture and change our understanding of approach to singularity in the classical theory is a promising arena. Interesting results in this direction have started appearing, including on singularity resolution in Bianchi models \cite{madrid-bianchi,lsubianchi} and quantum Kasner transitions across bounces and selection rules on possible structures near the to be classical singularities \cite{gs2}.  

Finally, we note that sometimes the limitations of LQC have been used to shed doubts on its results. These arguments, mainly articulated by the authors of \cite{bojowald-paly}, claim that a fully covariant approach with validity beyond symmetry reduced scenarios produces physical results inequivalent to those obtained from LQC. In particular, it is argued that, in presence of inhomogeneities, there is an unavoidable change of signature, from Lorentzian to Euclidean, in an effective theory. The authors of this chapter disagree with the conclusions reached in \cite{bojowald-paly} and subsequent papers along these lines. It our view, although the conceptual points raised by those authors are indeed interesting, their analysis relies in a series of assumptions and approximations that make their results far from being conclusive. Furthermore, recent results on the validity of the effective theories show that care must be taken in generalizing certain conclusions from the effective description to the full quantum theory \cite{dgs2,squeezed}.

\section*{Acknowledgments}

We are grateful to Abhay Ashtekar, Aurelien Barau, Chris Beetle, Boris Bolliet, B. Bonga, Alejandro Corichi, David Craig, Peter Diener, Jonathan Engel, Rodolfo Gambini, Julien Grain, Brajesh Gupt, Anton Joe, Wojciech Kaminski, Alok Laddha, Jerzy Lewandowski,  Esteban Mato, Miguel Megevand, Jose Navarro-Salas,  William Nelson, Javier Olmedo, Leonard Parker, Tomasz Pawlowski, Jorge Pullin, Sahil Saini, David Sloan,
Victor Taveras, Kevin Vandersloot, Madhavan Varadarajan, Sreenath Vijayakumar, and Edward Wilson-Ewing for many stimulating discussions and insights. This work is supported in part by NSF grants PHY1068743, PHY1403943,
PHY1404240, PHY1454832 and PHY-1552603. This work is also supported by a grant from John Templeton Foundation. The opinions expressed in this publication are those of authors and do not necessarily reflect the views of John Templeton Foundation.

\end{document}